\documentclass[12pt]{article}
\usepackage[utf8]{inputenc}
\usepackage[T1]{fontenc}

\usepackage[compress]{cite}
\usepackage{IEEEtrantools,stackengine}
\stackMath

\usepackage{authblk}
\usepackage{fullpage}
\usepackage{amssymb,amsmath}
\usepackage{adjustbox}
\usepackage{graphicx}
\usepackage{floatpag}
\usepackage{float}
\usepackage{longtable}

\usepackage{siunitx}
\usepackage{csquotes}

\usepackage{xspace}

\usepackage{booktabs}
\usepackage{longtable}
\usepackage{multirow}
\usepackage[percent]{overpic}
\usepackage{caption}

\usepackage[dvipsnames, table]{xcolor}

\definecolor{darkgreen}{rgb}{0.0, 0.5, 0.0}

\definecolor{lightyellow}{HTML}{FFE699}
\definecolor{red_revision}{HTML}{FF0000}

\usepackage{setspace}
\usepackage[unicode=true]{hyperref}
\hypersetup{breaklinks=true,
            pdfborder={0 0 0},
            allcolors=black}
\usepackage[%
  sort&compress
]{cleveref}
  \Crefname{appendix}{Supplement}{Supplements}
  \Crefname{figure}{Fig.}{Fig.}

\usepackage{times}
\usepackage{enumerate}
\usepackage{enumitem}

\usepackage{subfig}
\usepackage{graphicx}

\usepackage{rotating}
\usepackage{tabularx}
\usepackage{dcolumn}
\usepackage{pdflscape}
\usepackage{rotating}
\usepackage{array}

\usepackage{titlesec}

\titleformat{\subsubsection}
  {\normalfont\itshape}{\thesubsubsection}{1em}{}

\titleformat{\subsection}
  {\normalfont\bfseries}{\thesubsection}{1em}{}

\renewcommand{\arraystretch}{1.2}

\allowdisplaybreaks

\usepackage{eurosym}
\usepackage{comment}

\usepackage{ntheorem}
\theoremseparator{:}

\usepackage{makecell}

\begin{document}


\title{\centering\LARGE\singlespacing 
Hyperlocal monitoring of urban activity reveals responses to heat exposure}

\renewcommand\Affilfont{\fontsize{9}{10.8}\selectfont}

\author[,1,2]{Dominique Geissler}
\author[3,4]{Felix Creutzig}
\author[5,6]{Ramit Debnath}
\author[1,2]{Stefan Feuerriegel\thanks{Corresponding author: feuerriegel@lmu.de}}

\affil[1]{LMU Munich, Munich, Germany}
\affil[2]{Munich Center for Machine Learning (MCML), Munich, Germany}
\affil[3]{Potsdam Institute for Climate Impact Research, Potsdam, Germany}
\affil[4]{University of Sussex, Falmer, United Kingdom}
\affil[5]{University of Cambridge, Cambridge, United Kingdom}
\affil[6]{California Institute of Technology, Pasadena, United States}

\date{}

\maketitle

\newpage

\begin{abstract}\normalfont
\noindent
Rising temperatures create new challenges for local heat adaptation. Yet, it remains unclear how urban activity changes during hot periods and which urban environments people concentrate in as temperatures rise. Here, we perform a hyperlocal spatiotemporal analysis of urban activity across 10 German cities over a two-month period in 2024 with different levels of heat exposure. To monitor urban activity, we use fine-grained telecommunication data to map locations of people with high spatio-temporal resolution (i.e., hourly at $100\text{m} \times 155\text{m}$ grid cells), yielding more than 100 million data points. We then link activity counts with hourly weather records and point-of-interest data. We find that sustained periods of hot weather, defined as at least three consecutive days with daily maximum temperatures $\geq 25$°C, are characterized by below-expected city-wide presence, with activity counts that are 1.5 percentage points below regular urban activity. During hot periods, urban activity concentrates more strongly around leisure- and culture-oriented amenities (e.g., cafés or swimming pools), with an increase of up to around 10 percentage points relative to cooler days, while public-service environments (e.g., educational and health facilities) show weaker or negative shifts. Our study provides policy-makers with fine-grained monitoring of which urban areas attract citizens during heat exposure, which can enable evidence-based, spatially-targeted urban heat adaptation plans.
\end{abstract}

\flushbottom
\maketitle
\thispagestyle{empty}

\sloppy
\raggedbottom


\newpage
\section{Main}
\label{sec:introduction}

Cities are increasingly affected by climate change \cite{Lwasa.2022, Creutzig.2025}. Rising temperatures and more frequent heat can harm people's health \cite{Ebi.2021, Rouse.2026}, while urbanization further amplifies local heat exposure through the urban heat island (UHI) effect \cite{Milan.2015, Mohajerani.2017}, which is shaped by city size \cite{Oke.1973} and land-use \cite{Alavipanah.2015}. Urban heat exposure, however, depends not only on where high temperatures occur but also on where people live and where they move during hot periods \cite{Tuholske.2021, Yin.2021}. Hence, accurate information on where people are present during hot periods is needed to support urban adaptation planning.

The Intergovernmental Panel on Climate Change (IPCC) highlights cities as key sites where climate risks, infrastructure, land use, governance, and social vulnerability interact \cite{Lwasa.2022, ClimateADAPT.2025, Bai.2023, Creutzig.2025}, making cities important intervention points for climate adaptation \cite{McCormick.2013, Hsu.2020}. Several policy frameworks, including the European Climate Law \cite{EuropeanClimateLaw.2021} and the German Climate Adaptation Law \cite{KAnG.20.12.2023}, now require local governments to develop adaptation plans and conduct local risk analyses. Yet, such planning often lacks fine-grained evidence on actual urban activity during heat exposure. Static maps of temperature, land cover, or residential population provide important information, but they do not reveal where people are present during the day as temperatures change or which urban environments they use less, and, therefore, where targeted measures such as cooling, shading, drinking water access, or public-health communication may be most relevant.

Urban heat exposure depends on behavioral responses to weather conditions. Heat can change the timing of mobility \cite{Tian.2024, Batur.2024}, reduce mobility \cite{Stechemesser.2023, Renninger.2026}, alter the use of public transport \cite{Stechemesser.2023}, and affect the use of public and semi-public spaces \cite{Foshag.2020, Derakhshan.2023}. These responses are likely to differ across urban environments \cite{Wang.2025, Creutzig.2015}. During hot weather, people may reduce some forms of activity while concentrating around places that provide leisure, social interaction, shade, or green space. The availability and distribution of such environments are partly captured by points-of-interest (POI), such as green spaces, services, retail, food, culture, and other public-facing infrastructure. Such amenities drive urban vibrancy \cite{Botta.2021, Tu.2020} and help explain the spatial distribution of urban activity. Here, we analyze how the spatiotemporal patterns of urban activity changes in response to heat exposure.

Existing evidence on such behavioral exposure remains limited. Surveys and time-use data provide valuable insights into how heat changes daily routines \cite{Batur.2024}, but they often cover only selected locations, selected user groups, or short time windows. Transport system data capture adaptation in well-defined settings, such as subway use \cite{Stechemesser.2023, Zhao.2026}, while leaving activities outside the transportation network unobserved. Mobile phone data offer a complementary approach to measuring urban activity at larger spatial and temporal scales, and often cover a broader share of the population. However, many studies using mobile phone data analyze general urban mobility or activity patterns \cite{Deville.2014, Nachtigall.2025, Persson.2021}, but do not focus on the effects of heat exposure. Other studies use telecommunication data to analyze heat-related travel between cities \cite{Renninger.2026} or heat exposure inequality within cities \cite{DuranSala.2026, Tian.2024, Derakhshan.2023, Cai.2026}. Another stream uses telecommunication data to compute heat risk indices \cite{Holec.2021, Derakhshan.2025, Yang.2025}, but without analyzing the effect of heat on urban activity. So far, fine-grained spatio-temporal evidence on how urban activity changes during heat exposure is missing, but is needed to inform urban heat adaptation plans.

Here, we perform a spatiotemporal analysis of urban activity across 10 German cities (covering around 11.8 million people) during heat exposure. To monitor urban activity, we combine high-resolution telecommunication data over a two-month summer period in 2024 that measure hourly activity counts at the level of $100\text{m} \times 155\text{m}$ grid cells, yielding more than 100 million observations. We link these activity counts to hourly weather observations and POI data to analyze how within-city activity changes during hot periods, defined as days with daily maximum temperatures $\geq 25^\circ$C. Using 126,000 POIs across six categories, we examine how activity changes across different urban environments. Germany provides an informative empirical setting because cities have dense urban activity, varying urban form, and a policy context in which local heat adaptation planning is increasingly required \cite{KAnG.20.12.2023}.

Our study demonstrates the potential value of location-inferred telecommunication data as a scalable monitoring tool for urban heat adaptation. The results can directly support fine-grained heat adaptation planning by identifying locations and types where people are present during hot periods and which urban environments continue to attract activity; these areas may therefore require targeted measures \cite{Kopp.2026} such as shading, cooling infrastructure, drinking-water access, public-space redesign, or heat-health communication. In this way, high-resolution activity data can inform evidence-based, spatially-targeted approaches to urban heat adaptation.

\clearpage

\begin{figure}
    \centering
    \includegraphics[width=0.6\linewidth]{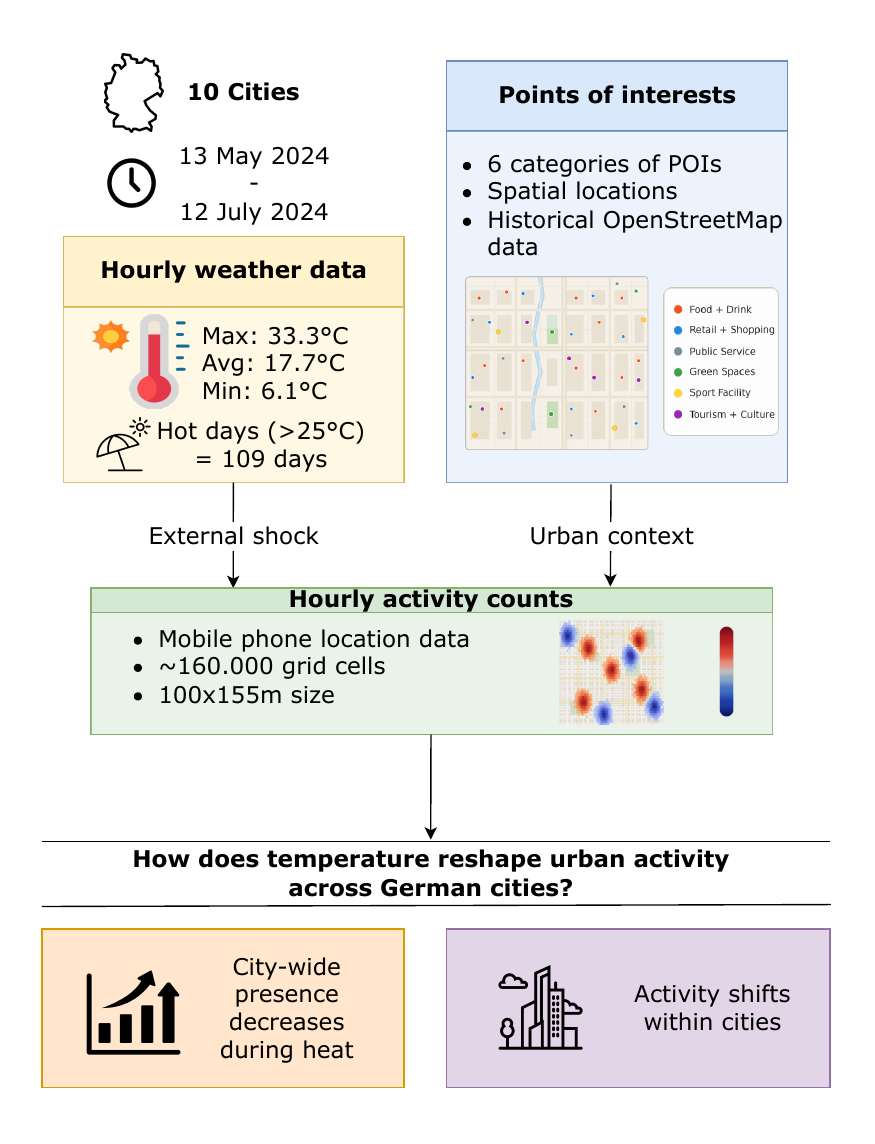}
    \caption{\textbf{Monitoring urban activity from telecommunication data during heat exposure in cities.} To link heat, urban context, and observed activity in cities, our study combines hourly weather observations, historical OpenStreetMap points-of-interest data, and anonymized and aggregated hourly counts of detected mobile devices across 10 German cities from 13 May to 12 July 2024. The telecommunication data measure observed activity in 100m $\times$ 155m grid cells, covering more than 100 million activity counts across $\sim$160,000 grid cells. Hot days are defined as days with daily maximum temperature $\geq 25^\circ$C. Weather is treated as the external temperature stressor, while points-of-interest (e.g., cafés, swimming pools) describe the urban context in which activity takes place. By linking these data at high spatial and temporal resolution, we analyze how heat exposure is associated with changes in city-wide presence and in the spatial distribution of activity within cities.}

    \label{fig:overview_data}
\end{figure}

\clearpage

\section{Results}
\label{sec:results}

\subsection{Monitoring human activity from telecommunication data in response to heat exposure}

\begin{figure}
\thisfloatpagestyle{empty}
\centering

\begin{minipage}[t]{0.25\textwidth}
    \centering
    \begin{overpic}[width=\linewidth]{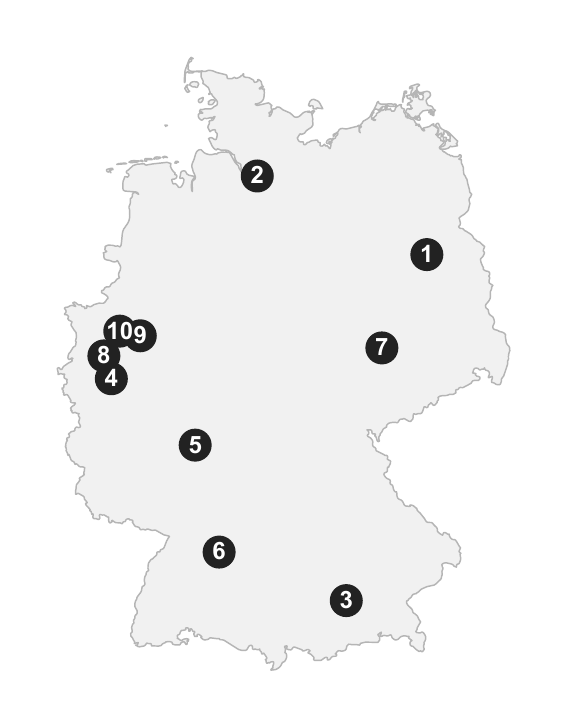} 
    \put(0,92){\textbf{a}} \end{overpic}
\end{minipage}
\begin{minipage}[t]{0.45\textwidth}
    \centering
    \begin{overpic}[width=\linewidth]{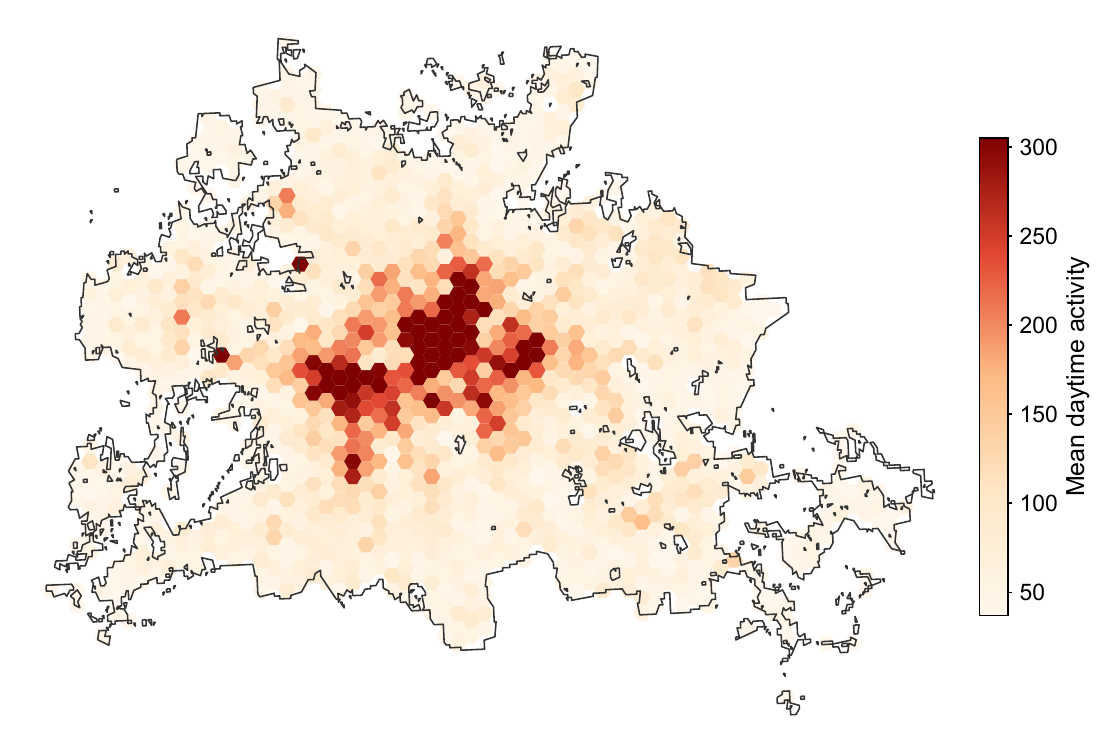}
    \put(0,66){\textbf{b}} \end{overpic}
\end{minipage}

\centering
\begin{minipage}[t]{0.8\textwidth}
    \centering
    \begin{overpic}[width=\linewidth]{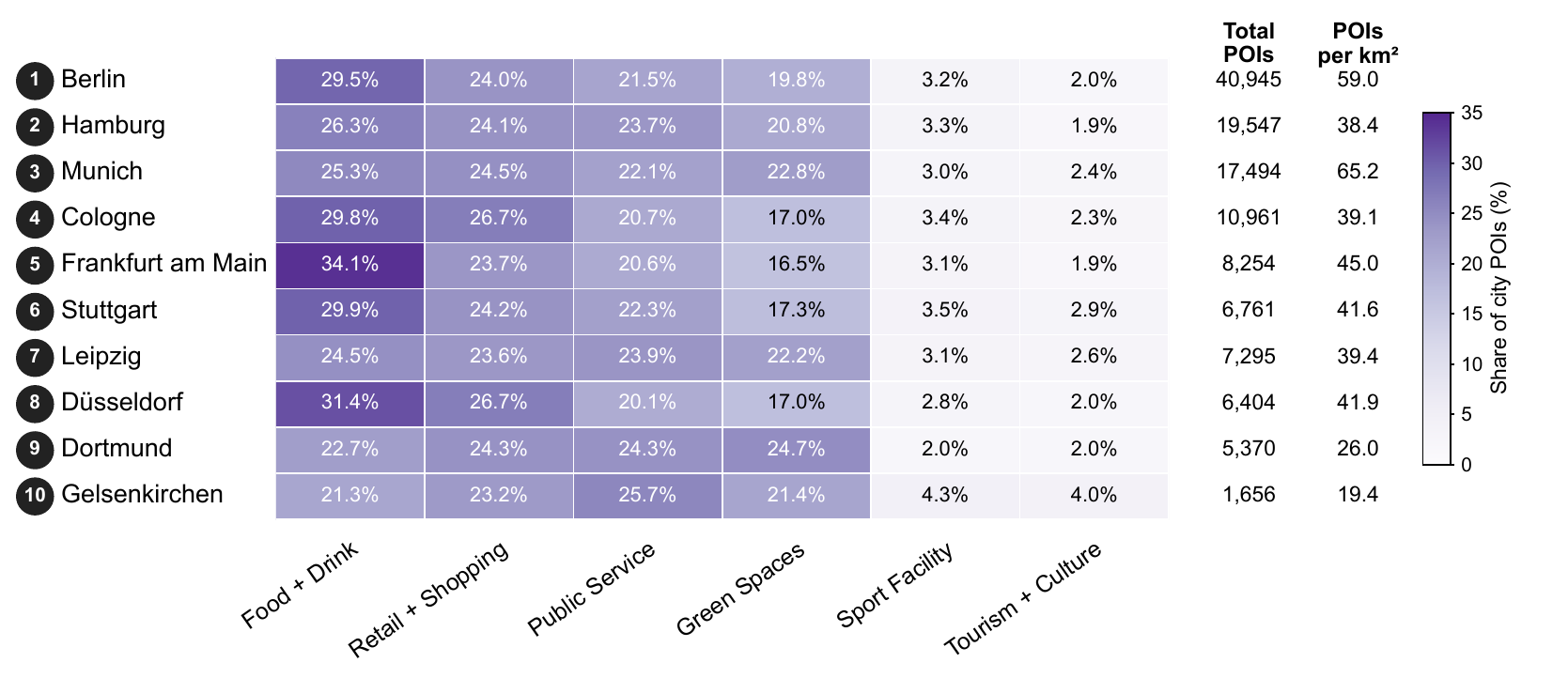}
    \put(7,42){\textbf{c}} \end{overpic}
\end{minipage}

\centering
\begin{minipage}[t]{0.5\textwidth}
    \centering
    \begin{overpic}[width=\linewidth]{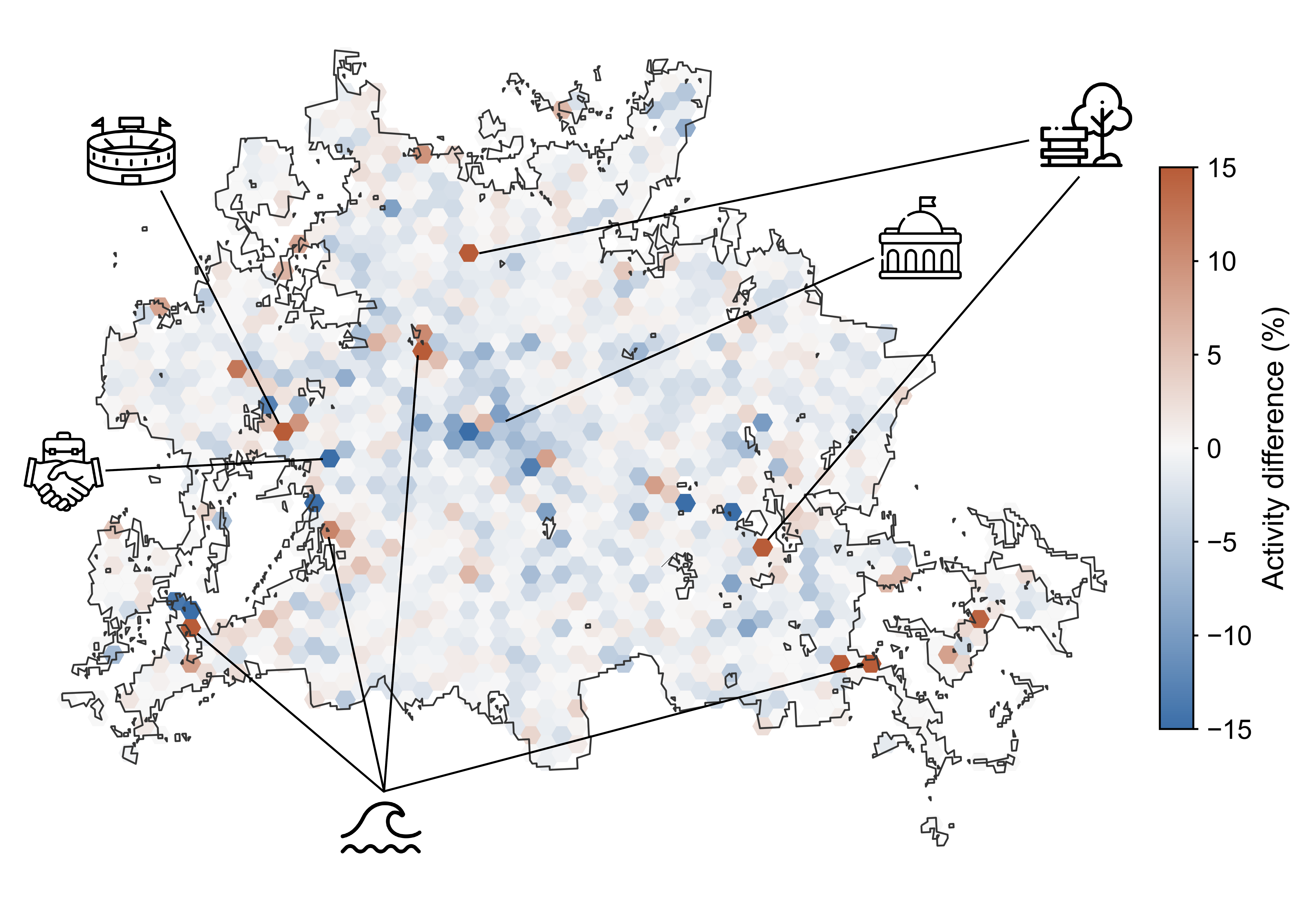}
    \put(0,63){\textbf{d}} \end{overpic}
\end{minipage}

\caption{
\textbf{Spatial overview of summer urban activity and urban context across German cities from 13 May to 12 July 2024.}
\textbf{a},~Location of the ten cities in Germany. Numbered markers correspond to the cities listed in panel \textbf{c}.  
\textbf{b},~Average daytime activity in Berlin. The fine-grained 100 m $\times$ 155 m cells were aggregated to hexagons of approximately 900 m for better visualization. 
\textbf{c},~Composition of points-of-interest by city. The two right columns report the total number of POIs and the number of POIs per km$^2$. Details about the aggregation of OpenStreetMap tags into POI categories are reported in Supplementary Table~\ref{tab:poi_categories}.
\textbf{d},~Observed percentage difference in daytime activity on hot days, defined as days with daily maximum temperature $\geq 25\,^{\circ}$C, relative to cooler days below $25\,^{\circ}$C, using the same spatial aggregation. Hexagons were aggregated to 900 m for better visualization. Red cells indicate areas with relatively higher activity on hot days, while blue cells indicate areas with relatively higher activity on cooler days. The annotations highlight selected urban contexts to aid interpretation: sports and event infrastructure, business and commercial areas, civic and cultural destinations, parks and green spaces, and waterfront locations such as rivers and lakes.
Similar plots for all other cities are reported in Supplementary Figure~\ref{fig:activity_change_maps_all_cities}.
}
\label{fig:spatial_overview}
\end{figure}

To analyze the response to heat exposure, we collected large-scale urban activity data from 10 German cities over a two-month period from 13 May to 12 July 2024. Together, these cities account for 11.76 million residents, and range from Berlin with 3.76 million residents to Gelsenkirchen with 0.26 million residents (Figure~\ref{fig:spatial_overview}\textbf{a}; Supplementary~Table~\ref{tab:city_population}). For this, hourly human movements derived from telecommunication data were obtained from a major telecommunications provider in Germany (see Section~\nameref{sec:methods}). The telecommunications provider anonymized and aggregated the data before transfer, and the research team only accessed hourly counts at the grid-cell level, without individual identifiers, subscriber information, or individual movement trajectories. All data processing followed privacy-preserving procedures in line with the European General Data Protection Regulation (GDPR).  Overall, the data contain more than 100 million anonymized and aggregated counts of people on 100m $\times$ 155m grid cells for $\sim$160,000 grid cells in total. 

To describe temperature-related change in urban activity, we also collected hourly weather observations from Deutscher Wetterdienst (DWD; German Meteorological Service) for each city \cite{DWD.2024}, with temperatures ranging from 6.1°C to 33.3°C over the two-month period. We define hot days as days with a daily maximum temperature $\geq 25,^\circ$C, following the DWD definition of a summer day \cite{DWD.2026.Sommertag}. During the observation window, there are 109 observations with hot days across the 10 cities. We further link the activity data to $\sim$126,000 POI from OpenStreetMap (Figure~\ref{fig:spatial_overview}\textbf{b}). We aggregated the POI tags from OpenStreetMap into six POI categories, namely, covering food and drink, retail and shopping, public services, green spaces, sport facilities, and tourism and culture (details in Supplementary Table~\ref{tab:poi_categories}).

Average activity per grid cell (counted in number of persons per hour) differs substantially across cities, ranging from 69.52 in Gelsenkirchen to 131.72 in Frankfurt am Main, and is higher on workdays than on non-workdays in all cities. The spatial distribution of daytime activity is illustrated for Berlin as an example (Figure~\ref{fig:spatial_overview}\textbf{c}, with activity aggregated to approximately 900 m hexagons for readability). We further observe large percentage differences in daytime activity on hot days relative to cooler days below 25°C (Figure~\ref{fig:spatial_overview}\textbf{d}). Similar plots for all other cities are reported in Supplementary Figure~\ref{fig:activity_change_maps_all_cities}.

\clearpage

\subsection{Heat exposure impacts whether people remain in urban areas}
\label{subsec:city-wide_presence}

Heat exposure may alter urban activity because high temperatures reduce thermal comfort, particularly in densely built environments \cite{Nikolopoulou.2003, Lin.2009}. This mechanism is consistent with published evidence that heat changes mobility, daily routines, and time use, including reductions or temporal shifts in travel and outdoor activity during hot conditions \cite{Stechemesser.2023, Tian.2024, Batur.2024, Renninger.2026}. One possible response is that people spend less time in central urban areas and shift their activity to places outside of cities, such as lakes, forests, or suburban green spaces, for which the cooling potential is well documented \cite{Bowler.2010, Norton.2015, Gunawardena.2017}. In our data, such behavior should appear as a decline in observed presence within the city because activity outside the city boundaries is no longer recorded.

\begin{figure}
    \centering
    \begin{overpic}[width=\linewidth]{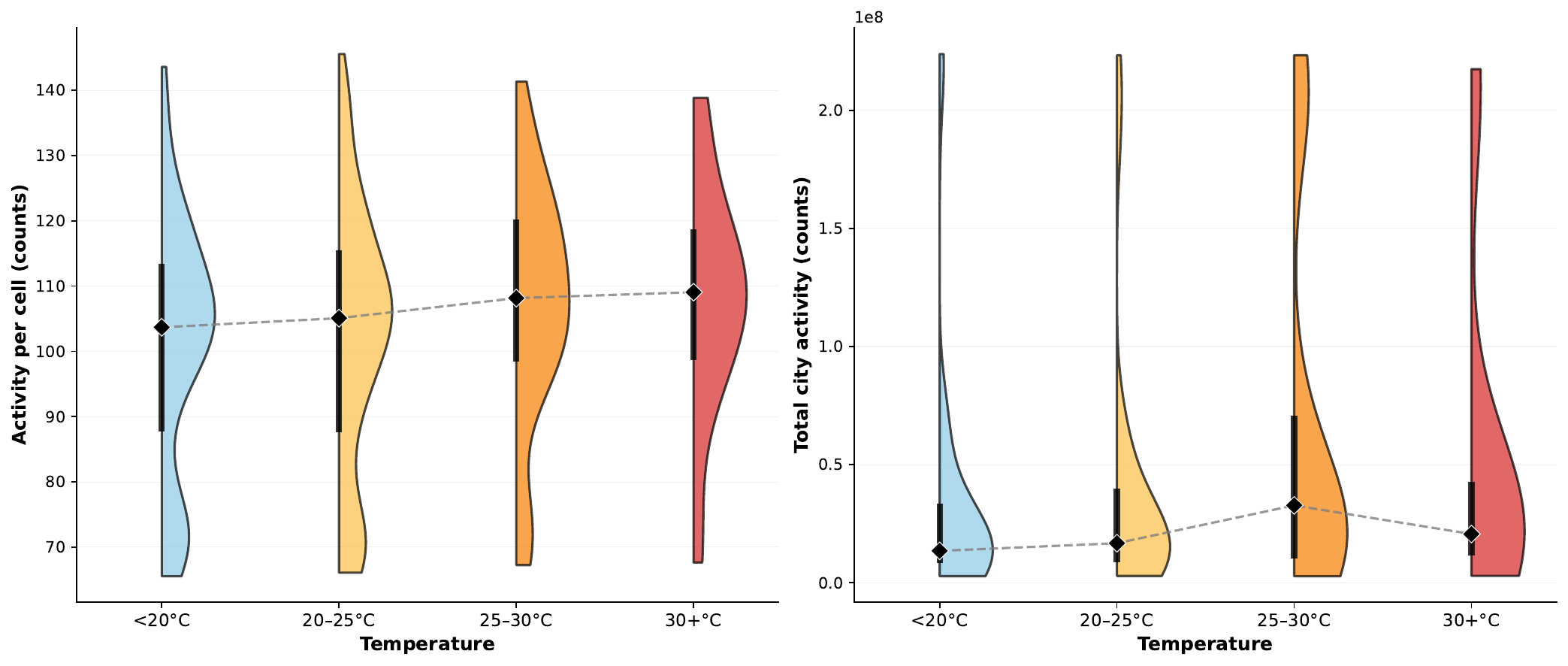} 
    \put(0,42){\textbf{\textcolor{black}a}} 
    \put(52,42){\textbf{\textcolor{black}b}} 
    \end{overpic}
    \caption{\textbf{Daily urban activity varies non-linearly with maximum temperature.} Half-violin plots show the distribution of daily urban activity across all cities by daily maximum temperature bin. \textbf{a}, Mean activity per grid cell. \textbf{b}, Total city activity. Temperature bins are defined by daily maximum temperature in $^\circ$C. Diamonds represent the median activity, and thick whiskers represent the interquartile range.}
    \label{fig:daily_activity_all_city_temp}
\end{figure}

To assess this relationship empirically, we examine the city-wide presence as daily total city activity and as average per grid cell activity during varying temperatures. The observed patterns indicate a non-linear relationship between daily maximum temperature and urban activity. Across all cities, observed activity increases with daily maximum temperature until around $30,^\circ$C, before stagnating or declining at higher temperatures (see Figure~\ref{fig:daily_activity_all_city_temp}). Days with maximum temperatures between $25\,^\circ$C and $30\,^\circ$C show significantly higher activity than cooler days (average activity per grid cell: $<20\,^\circ$C vs. $25$--$30\,^\circ$C: $t=-2.579$, $p\leq0.05$; $20$--$25\,^\circ$C vs. $25$--$30\,^\circ$C: $t=-2.302$, $p\leq 0.05$; total city activity: $<20\,^\circ$C vs. $25$--$30\,^\circ$C: $t=-3.285$, $p\leq0.001$; $20$--$25\,^\circ$C vs. $25$--$30\,^\circ$C: $t=-2.817$, $p\leq 0.01$). By contrast, differences involving days above $30\,^\circ$C are not statistically significant at conventional levels, which may partly reflect the small number of very hot days in the observation period ($N=21$).
 
\begin{figure}
    \centering
    \includegraphics[width=0.95\linewidth]{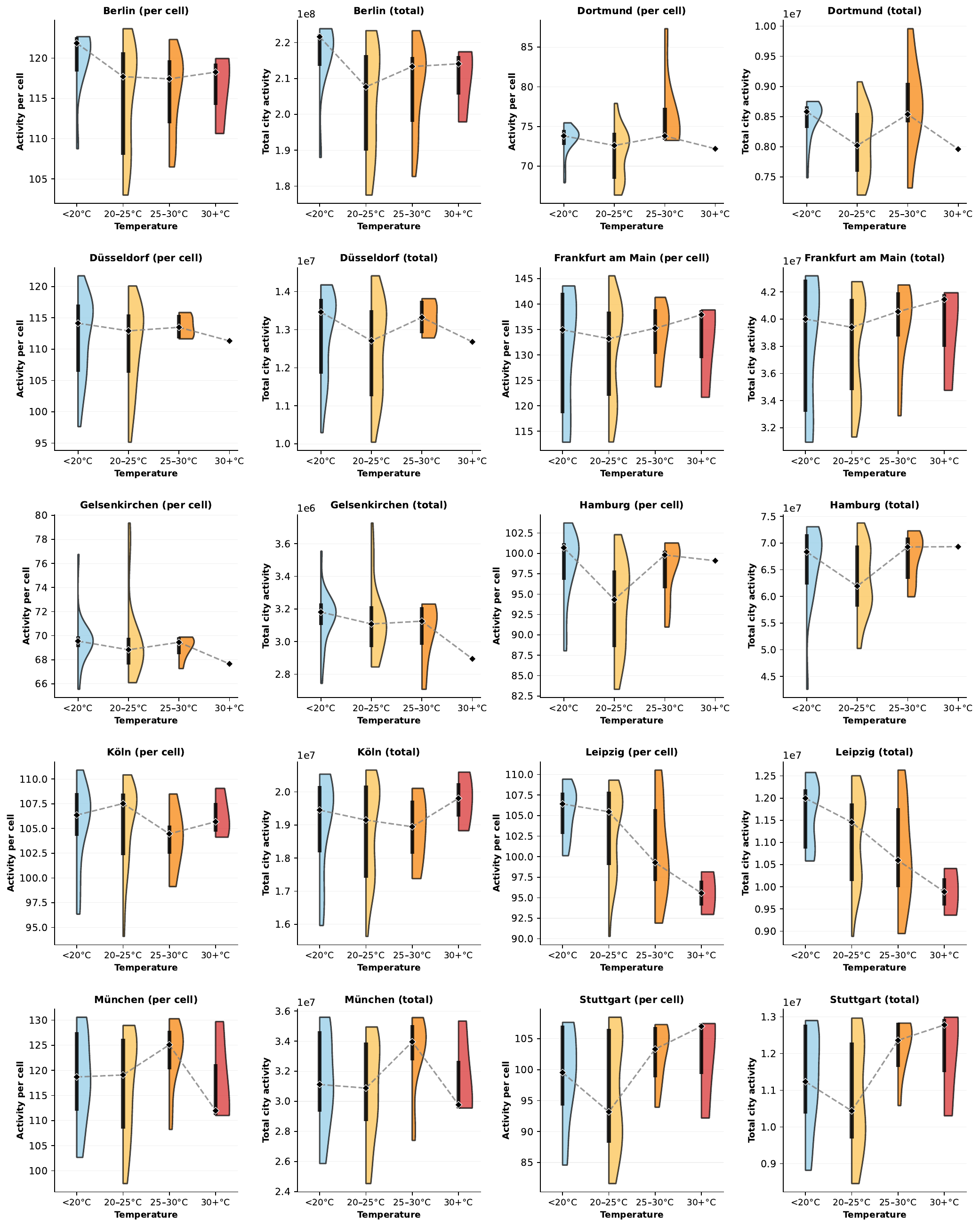}
    \caption{\textbf{Heterogeneity in daily urban activity across maximum daily temperature.} Half-violin plots show the distribution of daily urban activity by daily maximum temperature bin for each city, summarized as average activity per grid cell (left) and as total city activity (right). Diamonds represent the median activity, and black boxes represent the interquartile range.}
    \label{fig:daily_activity_per_city_temp}
\end{figure}

City-specific patterns are heterogeneous (Figure~\ref{fig:daily_activity_per_city_temp}). In most cities, observed presence is lower during mild temperatures (defined as days between $20\,^\circ$C and $25\,^\circ$C), and higher on both colder days ($< 20\,^\circ$C) and very hot days ($> 25\,^\circ$C). Further descriptive statistics, including city-specific medians and percentage differences, are reported in Supplementary Table~\ref{tab:temperature_bin_activity_compact_median}. A notable exception is Leipzig, where observed presence declines monotonically as daily maximum temperature increases.

To study whether temperature-related changes in urban activity differ between workdays and non-workdays, we regress city-wide presence on midday temperature separately by day type using ordinary least squares (Table~\ref{tab:temperature_activity_midday_daytype}). On workdays, urban activity is likely structured by commuting and work routines, which may limit temperature-related changes in mobility. Consistent with this interpretation, the pooled weekday estimate is positive but not statistically distinguishable from zero. At the city level, estimates vary across cities, with several negative coefficients, indicating heterogeneity in how weekday activity changes with midday temperature. In Berlin, Düsseldorf, Gelsenkirchen, and Leipzig, a one-degree Celsius increase is linked to approximately 0.6\% to 1.0\% lower urban activity, as measured by activity counts. On non-workdays, when activity is likely more discretionary, the estimated relationship is more consistently positive. Pooled across all cities, a one-degree Celsius increase in midday temperature is linked to approximately 6.1\% higher city-wide activity. City-specific estimates again show heterogeneity, with mostly positive coefficients but some deviations from this pattern. Overall, the results indicate that warmer midday temperatures are more strongly associated with discretionary non-workday activity than with regular weekday activity.

\begin{table}[!htbp]
\centering
\small
\begin{tabular}{@{}lrrr@{\hspace{.5cm}}rrr@{}}
\toprule
 & \multicolumn{3}{c}{Workday} & \multicolumn{3}{c}{Non-workday} \\
\cmidrule(lr){2-4} \cmidrule(lr){5-7}
City & Coef. & Std. error & $p$-value & Coef. & Std. error & $p$-value \\
\midrule
All cities & 0.0228 & 0.0144 & 0.113 & 0.0594 & 0.0242 & $\leq 0.05$ \\
Berlin & --0.0060 & 0.0023 & $\leq$ 0.01 & 0.0057 & 0.0032 & 0.080 \\
Dortmund & --0.0073 & 0.0047 & 0.118 & 0.0052 & 0.0072 & 0.469 \\
Düsseldorf & --0.0070 & 0.0033 & $\leq 0.05$& 0.0080 & 0.0084 & 0.340 \\
Frankfurt am Main & --0.0050 & 0.0035 & 0.146 & 0.0096 & 0.0038 & $\leq 0.05$ \\
Gelsenkirchen & --0.0096 & 0.0044 & $\leq 0.05$ & 0.0055 & 0.0038 & 0.151 \\
Hamburg & --0.0006 & 0.0062 & 0.920 & 0.0077 & 0.0053 & 0.148 \\
Cologne & --0.0055 & 0.0034 & 0.103 & 0.0080 & 0.0049 & 0.100 \\
Leipzig & --0.0105 & 0.0043 & $\leq 0.05$ & --0.0100 & 0.0038 & $\leq 0.01$ \\
Munich & 0.0028 & 0.0022 & 0.213 & 0.0034 & 0.0025 & 0.174 \\
Stuttgart & 0.0011 & 0.0032 & 0.718 & 0.0034 & 0.0041 & 0.405 \\
\bottomrule
\end{tabular}
\caption{\textbf{Estimated relationship between midday temperature and city-wide activity by city and day type.} The table reports ordinary least squares estimates from regressions of logged city-wide activity on midday temperature, estimated separately by city and day type. Because the outcome is logged, coefficients can be interpreted approximately as percentage changes in city-wide activity associated with a one-degree Celsius increase in midday temperature. For example, a coefficient of 0.059 corresponds to approximately 5.9\% higher city-wide activity. The pooled estimates indicate a stronger positive relationship on non-workdays than on workdays, while city-specific estimates show substantial heterogeneity. Results are robust to using raw activity counts as the outcome variable (Supplementary~Table~\ref{tab:temperature_activity_midday_daytype_raw}).}
\label{tab:temperature_activity_midday_daytype}
\end{table}

\subsection{City-wide activity during sustained hot periods}

Consecutive hot days can increase the risk because heat stress accumulates and physiological recovery becomes more limited \cite{Anderson.2009, WHO.Heat.2026}. We therefore analyze city-wide presence during sustained hot periods, with at least three consecutive days on which the daily maximum temperature is unusually high \cite{DWD.2026Hitzewelle}, which we define as at least three consecutive days with a daily maximum temperature $\geq 25^\circ$C, following the definition of the Deutscher Wetterdienst (DWD; German Meteorological Service) of a summer day \cite{DWD.2026.Sommertag}.

Our observation window includes a total of 19 sustained hot periods across the ten cities, corresponding to 89 days. For example, three sustained hot periods are each observed in Berlin, Hamburg, and Munich, respectively. Mean daily maximum temperatures during the hot periods range from $25.7^\circ$C in Berlin to $29.6^\circ$C in Hamburg and Frankfurt am Main. A detailed overview is provided in Supplementary~Table~\ref{tab:consecutive_hot_days}.

To estimate whether city-wide presence during sustained hot periods deviates from regular temporal patterns, we first model expected presence as a function of workday status, public holidays, school holidays, and a dummy for the UEFA EURO 2024 period. UEFA EURO 2024 was the 2024 European football championship hosted in Germany, which may have thus affected urban presence through match-day travel, public viewing, and tourism. We then compare observed city-wide presence with the model-predicted expected presence and express the ratio as an activity index. The index measures relative deviations from expected activity, where a value of 102 would indicate 2\% higher observed presence than expected under comparable calendar and event conditions, while a value of 98 would indicate 2\% lower observed presence.

\begin{figure}
    \centering
    \includegraphics[width=0.95\linewidth]{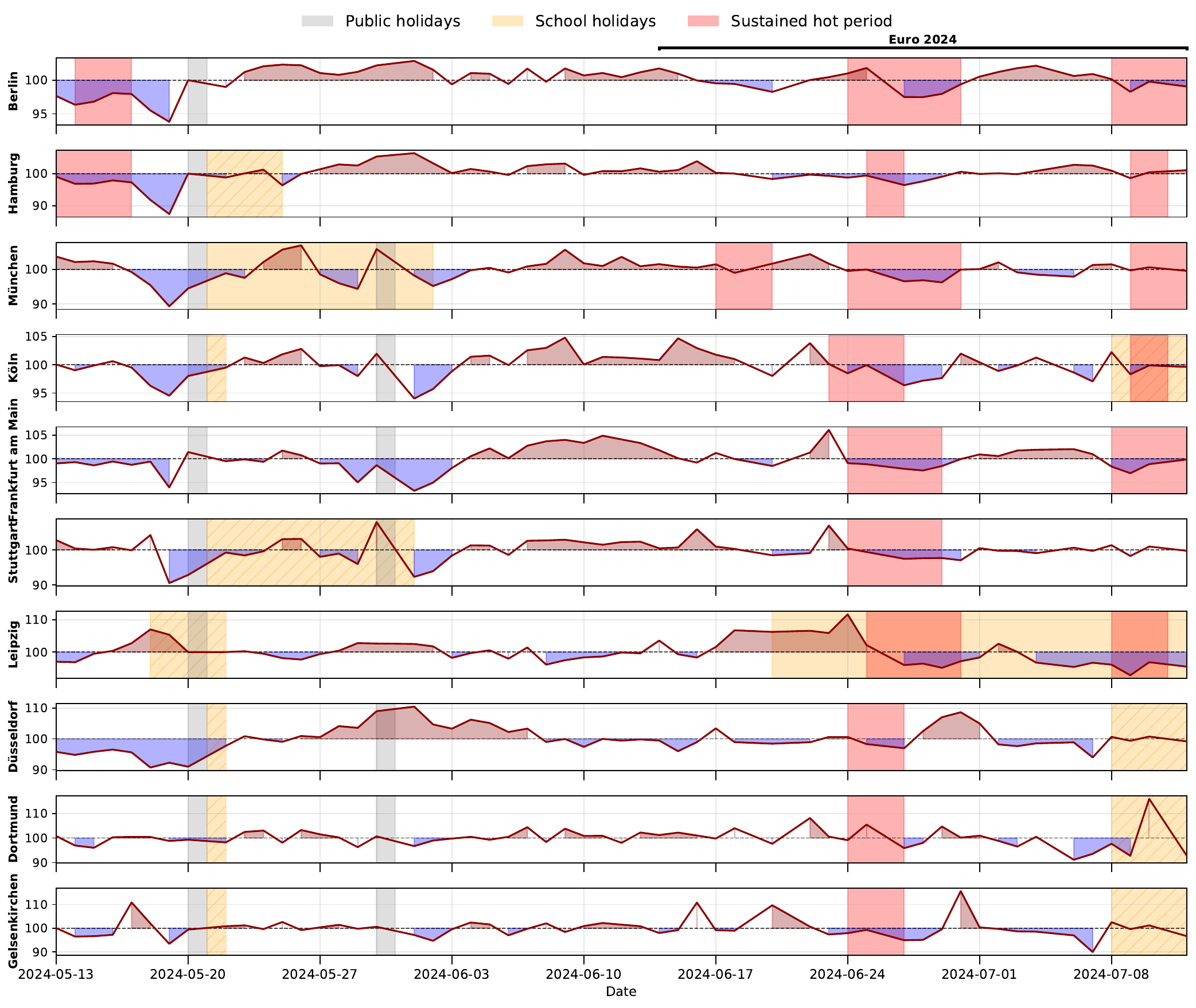}
    \caption{\textbf{Difference between observed and expected city-wide presence during sustained hot periods.}
    Expected presence is estimated from regular temporal patterns, adjusting for workday status, public holidays, school holidays, and a dummy for the UEFA EURO 2024 period. UEFA EURO 2024 was hosted in Germany (14 June to 14 July 2024) and is included because match-day travel, public viewing, and tourism may have affected city-wide presence independently of temperature. The figure reports an activity index defined as observed presence divided by model-predicted expected presence, multiplied by 100. The dark red line shows the activity index. Red areas indicate values above 100, while blue areas indicate values below 100. Values below 100 indicate fewer people observed within the city than expected ceteris paribus. In several cities, sustained hot periods, using the Deutscher Wetterdienst (DWD; German Meteorological Service) definition of a summer day with $\geq25^\circ$C, coincide with below-expected presence after adjusting for public holidays, school holidays, and the Euro 2024 period. The pattern suggests that persistent heat may be associated with reduced city-wide presence.}
    \label{fig:heatwaves_activity}
\end{figure}

Observed city-wide presence often falls below expected levels during sustained hot periods (see Figure~\ref{fig:heatwaves_activity}). On average, the activity index is 100.23 on mild days ($<25^\circ$C),  and 98.45 during sustained hot periods (using the above definition from Deutscher Wetterdienst). The deviation from expected presence is statistically significant for sustained hot periods ($t=-6.73$, $p<0.001$), with 60 hot-period days below the expected value. Together, these results indicate that sustained hot periods are associated with a modest but systematic reduction in city-wide presence relative to expected urban activity.

\subsection{Temperature-dependent activity around points-of-interest}
\label{subsec:poi_redistribution}

Hotter conditions may change which urban environments people use; for example, by shifting activity toward leisure, retail, or green space amenities. Such within-city characteristics are often captured through POIs \cite{Ganter.2022, Botta.2021, Tu.2020, Kadar.2018}, which provide spatial information on the types and functions of urban locations. Such amenities drive urban vibrancy \cite{Botta.2021, Tu.2020} and help explain where activity concentrates within cities. Especially so-called third spaces (i.e., cafés, parks, libraries, and cultural venues) where people spend time outside home and work \cite{Botta.2021} are particularly relevant in the context of urban heat because they are discretionary destinations that people may seek or avoid depending on weather conditions \cite{Zhao.2026}. 

To examine whether activity around these POI-characterized environments changes with temperature, we estimate fixed-effects regressions at the grid-cell-hour level. The models relate activity counts to temperature bins, POI category indicators, and their interactions, while accounting for regular temporal patterns and city-wide activity levels. The interaction terms capture whether activity in grid cells characterized by a given POI category is higher or lower in a temperature bin relative to the reference temperature category and relative to otherwise comparable grid cells without that POI category. Here, we use six aggregated POI categories: \emph{Food \& Drink} (e.g., restaurants, cafés, bars), \emph{Retail \& Shopping} (e.g., shops, supermarkets, malls), \emph{Public Service} (e.g., administrative offices, schools, healthcare facilities), \emph{Green Spaces} (e.g., parks and playgrounds), \emph{Sport Facility} (e.g., sports centers, fitness centers, swimming pools), and \emph{Tourism \& Culture} (e.g., museums, galleries, attractions, stadiums). The detailed mapping of OpenStreetMap tags to these categories is provided in Supplementary Table~\ref{tab:poi_categories}.

\begin{figure}
    \centering
    \includegraphics[width=0.7\linewidth]{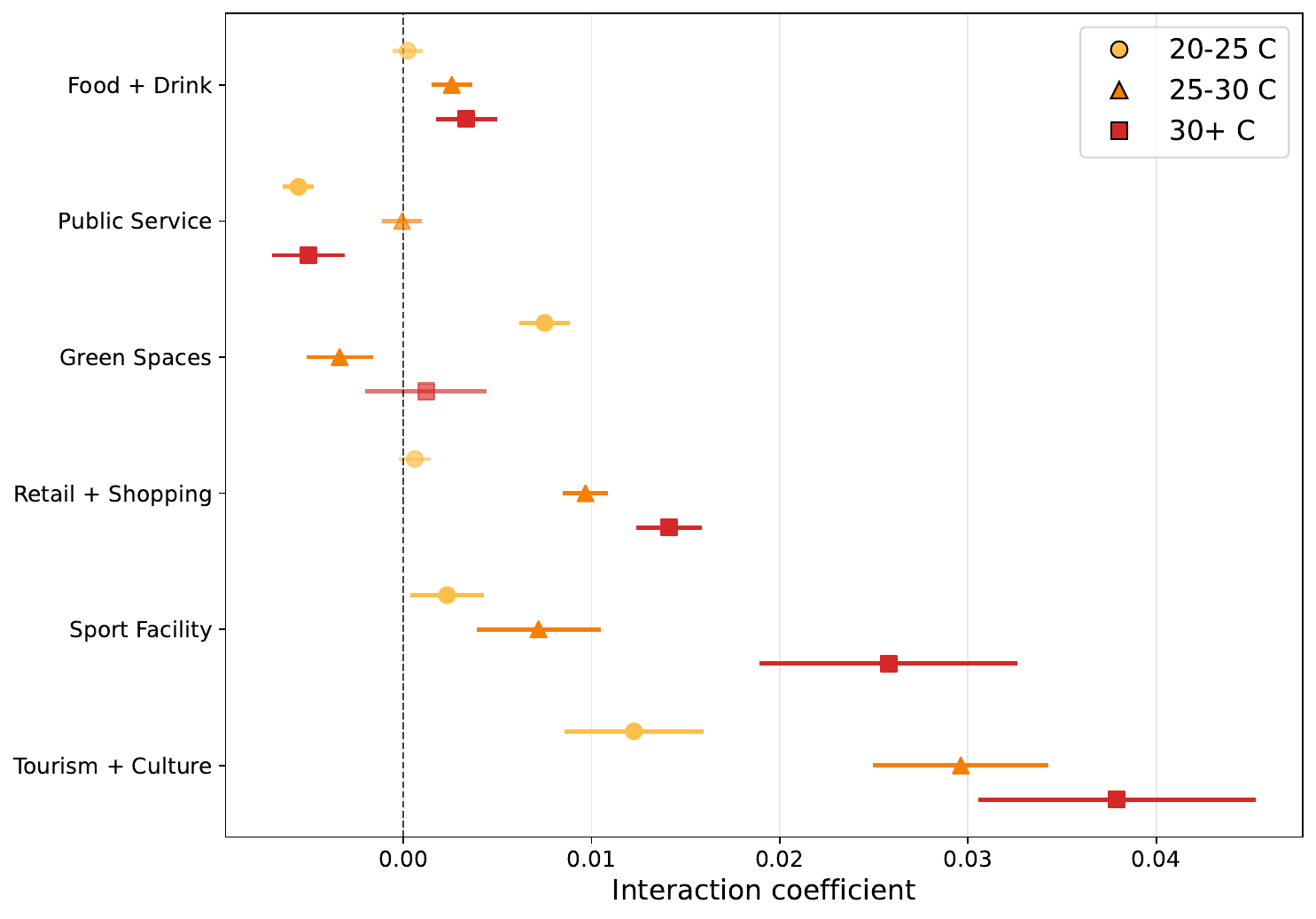}
    \caption{\textbf{Temperature-dependent differences in activity around POIs.} The figure reports estimated interaction coefficients from an across-city pooled regression linking temperature bins to activity around six POI categories: \emph{Food + Drink}, \emph{Retail + Shopping}, \emph{Public Service}, \emph{Green Spaces}, \emph{Sport Facility}, and \emph{Tourism + Culture}. The model is estimated at the grid-cell-hour level and includes controls for regular temporal patterns and city-wide activity levels. Because the outcome is logged activity, coefficients can be interpreted approximately as percentage differences in activity. Coefficients indicate whether activity in grid cells characterized by a POI category is higher or lower in a given temperature bin relative to the reference temperature category and relative to otherwise comparable grid cells without that POI category. Points denote coefficient estimates and whiskers denote 95\% confidence intervals.}
    \label{fig:regression_poi_pooled}
\end{figure}

The pooled estimates indicate that temperature-dependent activity differs across POI categories (see Figure~\ref{fig:regression_poi_pooled}). Activity around POIs from the \emph{Tourism \& Culture} increases most consistently in hotter temperature bins with up to 3.86\% for days with temperature $\geq 30^\circ$C, suggesting that tourism- and culture-related environments attract relatively more activity under warmer conditions. Activity around POIs from the categories \emph{Food \& Drink} and \emph{Retail \& Shopping} also tends to increase in hotter temperature bins (up to 0.35\% and 1.41\%, respectively), although the estimated magnitudes are smaller. By contrast, the estimates for \emph{Public Service} are weaker and often close to zero or negative, suggesting that hotter conditions are not associated with a comparable increase in activity around more functional service environments. The estimates for \emph{Green Spaces} and \emph{Sport Facility} are more mixed, indicating that these categories do not form uniform heat-attraction environments in the pooled analysis.

City-specific regressions show that these POI-related temperature patterns are heterogeneous across cities (see Supplementary Figure~\ref{fig:regression_poi}). For example, the positive pooled association for \emph{Tourism \& Culture} is more pronounced in large cities such as Berlin and Cologne than in others. This heterogeneity is further visible for \emph{Green Spaces}. In Munich, grid cells characterized by \emph{Green Spaces} are about 5\% higher in the hottest temperature bin than in the reference temperature category, relative to otherwise comparable grid cells and conditional on the model controls. By contrast, estimates for \emph{Green Spaces} are close to zero or negative in other cities, such as Hamburg. This suggests that green spaces are not uniformly associated with increased activity during hotter periods; their role likely depends on local accessibility, design, shade, and the specific types of green space captured in the POI data. This heterogeneity suggests that the role of environments characterized by POIs during hotter conditions depends on local urban structure, accessibility, and the specific types of amenities captured within each category.

\section{Discussion}
\label{sec:discussion}

This study analyzes how summer heat reshapes urban activity. Across ten German cities, observed city-wide presence follows a non-linear temperature pattern. Activity in the cities increased with warmer temperatures up to a daily maximum temperature of $30^\circ$C. For example, the total number of people in Munich increased by 10\% from $20$--$25^\circ$C to $25$--$30^\circ$C. At high daily maximum temperatures over $30^\circ$C, observed city-wide presence stagnated or declined. During sustained hot periods, observed city-wide presence falls systematically below expected levels. Across all cities, activity during sustained hot periods is on average 1.55\% lower than expected from regular calendar and event patterns. In addition to city-wide presence, heat exposure also influenced where activity is concentrated in the city. Activity increased around leisure, cultural, food, and retail amenities, while routine public-service environments showed weaker or negative shifts. 

These findings address an empirical gap in climate adaptation planning. Existing adaptation frameworks such as the IPCC report \cite{Bai.2023, Creutzig.2025}, the European Climate Law \cite{EuropeanClimateLaw.2021}, and Germany's Climate Adaptation Act \cite{KAnG.20.12.2023} require local governments to identify \emph{where} heat exposure arises in daily life and \emph{where} interventions should be targeted. Many heat risk assessments rely on temperature surfaces, residential population, or static vulnerability indicators \cite{Alavipanah.2015, Anderson.2009, Foshag.2020, Gunawardena.2017, Mohajerani.2017, Norton.2015, RuizMallen.2025}, which, while essential, do not fully capture everyday exposure because people move through the city for work, consumption, leisure, and social life. By linking heat, observed presence, and urban amenities, our study adds a behavioral layer to urban heat assessment, which can help identify not only where heat is physically intense but also the spatio-temporal patterns of exposure risk that arise as people move through urban environments during hot conditions.

Climate adaptation is shaped by local conditions because heat risks depend on built form, infrastructure, green and blue spaces, transport systems, and everyday patterns of urban activity\cite{Wang.2025, Creutzig.2015}. German cities provide a particularly informative setting for studying these local dynamics. They are dense, infrastructure-rich, and increasingly exposed to summer heat, with heat extremes projected to rise substantially over the coming decades even under moderate emissions scenarios \cite{WorldBank.2025}. At the same time, much of the existing evidence on urban heat and mobility has focused on cities in the United States or China, leaving Central European urban contexts comparatively less studied \cite{Duymus.2025}. Studying German cities, therefore, allows us to examine how heat exposure relates to everyday urban activity in a context where local adaptation planning is becoming increasingly urgent.

A key contribution of our study is to propose telecommunication-derived activity as a scalable monitoring system for urban heat adaptation planning. Compared with surveys, time-use diaries, or mobility traces \cite{Batur.2024, Zhao.2026, Stechemesser.2023}, monitoring based on aggregated telecommunication data offers four main strengths. First, it provides city-wide, high-resolution spatio-temporal information on observed presence. Second, it is not limited to self-reported routines, specific transport systems, or opt-in app users. Third, even though it is not perfectly representative, such monitoring likely captures a broader and more representative part of society because it includes all people carrying active SIM-enabled devices. Fourth, it can be implemented in an anonymized and aggregated form in a privacy-preserving and GDPR-compliant setting. These properties make telecommunication-derived activity monitoring particularly well suited for studying urban heat exposure across the spatial and temporal scales relevant for adaptation planning, including within-city variation, short-term changes over time, and fine-grained local activity patterns. 

The city-wide results show a non-linear relationship between temperature and observed urban activity. Across cities, observed activity tends to increase with warmer summer conditions up to around $30^\circ$C, before stagnating or declining at higher temperatures. Hence, moderate warmth is associated with higher urban presence, whereas higher temperatures may reduce presence within the city or shift activity to other locations. This pattern is consistent with the idea that moderate warmth can act as a pull factor for urban life in that people may spend more time in public, commercial, and recreational spaces when weather conditions are pleasant. At higher temperatures, however, heat may become a push factor that reduces outdoor exposure, changes daily routines, or shifts activity away from the city. This interpretation is consistent with earlier research showing that subway use in New York City follows a strong non-linear temperature response, with lower mobility at both cold and hot extremes \cite{Stechemesser.2023}, and with evidence that extreme heat reduces and restructures mobility across Spain \cite{Renninger.2026}. Sustained hot periods show a similar pattern more clearly, where observed presence falls below expected levels even after adjusting for calendar and event patterns. This aligns with evidence that heat can shift activity toward cooler morning and evening hours \cite{Tian.2024} and that people spend more time indoors during prolonged heat \cite{Batur.2024}.

The POI analysis shows that heat is associated not only with changes in overall urban presence, but also with changes in where activity is located within cities. Grid cells characterized by food and drink, retail, cultural, and sports amenities showed relatively higher activity as temperatures increased, whereas cells characterized by public-service functions showed weaker or no comparable increases. This suggests that activity during hotter conditions shifts more strongly toward discretionary urban environments than toward routine service environments. Public-service locations are often tied to fixed appointments or administrative tasks, so their use may be less responsive to temperature. By contrast, visits to food, retail, cultural, and recreational locations are more flexible in timing and location, and may therefore change more strongly with weather conditions \cite{Renninger.2026}. This interpretation is consistent with field evidence showing that the use of public urban spaces varies with thermal conditions \cite{Lin.2009, Nikolopoulou.2003}.

The stronger activity increases around food, drink, retail, and cultural amenities may reflect several features of these environments. Many of these locations are indoors or partly indoors and may provide cooler conditions than exposed outdoor spaces \cite{Dearman.2025}. They also offer services, consumption opportunities, and social activities, which give people a reason to spend time there during hot conditions. This is consistent with the idea of ``third places'', such as cafés, restaurants, shops, or cultural venues, as urban settings outside home and work where people meet and spend time \cite{Botta.2021, Gehl.2011}. Our results, therefore, suggest that the relationship between POI structure and urban activity depends partly on temperature. This extends earlier urban research linking POI structure to how ``vibrant'' cities are \cite{Botta.2021, Huang.2024}, and is consistent with evidence that semi-public indoor spaces, such as shopping centers, can function as informal cooling locations during heat \cite{Derakhshan.2023}.

The results for green spaces show substantial variation across cities. Activity around green spaces increased during hotter conditions in some cities, but changed little or declined in others. This suggests that green spaces do not function uniformly as heat refuges \cite{Yang.2025}. Their use during hot periods likely depends on local characteristics such as shade, water availability, accessibility, perceived safety, opening hours, and connections to surrounding neighborhoods \cite{Woo.2026}. This is consistent with evidence that the cooling effect of green spaces depends on vegetation type and tree composition \cite{Li.2024}. For adaptation planning, the presence of green space alone is therefore not sufficient, but policy-makers should carefully assess that the value during heat depends on whether people can reach and use it comfortably when temperatures are high \cite{Derakhshan.2025}.

These results can inform spatially-targeted adaptation policies. Environments that remain active under heat, such as cultural districts, shopping streets, food-and-drink clusters, and sports facilities, are relevant locations for shading, drinking-water provision, and temporary cooling \cite{Cai.2026}. Existing indoor or semi-public spaces with cooling, such as shopping centers or cultural venues \cite{Derakhshan.2025}, should be considered when planning local cooling infrastructure. For example, in Berlin, the Berliner Erfrischungskarte provides an example of such spatially explicit adaptation support by visualizing cool, windy, and shaded areas as well as infrastructure such as benches and drinking-water fountains that can support refreshment and lingering during hot weather \cite{Erfrischungskarte.2023}. Our activity-based evidence complements such cooling maps by showing where people are actually present during hot periods, and can therefore help prioritize which cooling opportunities may be most relevant in practice. Conversely, areas where activity declines during heat may indicate urban environments that become less usable under high temperatures and should be examined in local risk assessments\cite{Kopp.2026, Creutzig.2026}.

Our study is subject to several limitations. First, the telecommunication data measure aggregated presence per grid cell, not individual trajectories or visits to specific POI. The estimates, therefore, describe activity around amenity-rich cells rather than attendance at named venues. Second, the data are not perfectly representative. Counts depend on device ownership, whether people carry active SIM-enabled devices, and the technical processing of raw signals. Some groups, including children and older adults, may be less consistently captured. Counts may also be affected by behavioral differences in phone carrying, for example, when people leave phones behind during sports or other activities, and by multiple devices or SIM cards associated with the same person. Still, earlier research suggests that mobile phone data is highly effective in mapping population movements \cite{Deville.2014}. Third, the observation window covers two early-summer months in 2024. It includes sustained hot periods, but not late-summer heat or more extreme heat episodes. Fourth, we use station-based air temperature, which does not capture fine-scale differences in intra-urban microclimate. Finally, the analyses identify associations between temperature, urban context, and observed activity, rather than causal effects. 

Our study shows that urban heat exposure is shaped by where people spend time during hot conditions. Monitoring observed activity can help cities target adaptation measures to places where people are actually present, including active public, commercial, and leisure areas. More broadly, the framework contributes to a growing data-driven research agenda on urban heat adaptation, where machine learning and large-scale urban data are increasingly used to characterize heat exposure and support adaptation planning \cite{Hintz.2025}. At the same time, adaptation cannot replace mitigation; reducing future heat extremes remains essential \cite{Krayenhoff.2018}. Together, these findings support urban heat policy that combines climate mitigation with locally targeted adaptation based on how cities are used in practice.

\section{Methods}
\label{sec:methods}

\subsection{Telecommunication data}
\label{sec:telcom_data}
Telecommunication data were obtained from a German mobile network operator, aggregated to grids of 100m $\times$ 155m cells for ten German cities: Berlin, Cologne, Dortmund, Düsseldorf, Frankfurt am Main, Gelsenkirchen, Hamburg, Leipzig, Munich, and Stuttgart. The number of devices detected per grid cell was recorded at hourly intervals over a two-month period from 13 May to 12 July 2024. The data contain more than 100 million counts of people for $\sim$160,000 grid cells. All data were provided in anonymized and aggregated form in accordance with the European General Data Protection Regulation (GDPR). Ethical approval was obtained from the Institutional Review Board (IRB) at LMU Munich (Number ETH-SOM-081). 

Telecommunication-derived activity data were generated by the partner company from routine signal exchanges between SIM-enabled mobile devices and the mobile network from one mobile service provider (German D1 network). Mobile phones regularly connect to nearby antennas to maintain network service, producing technical metadata such as an anonymized device identifier, timestamp, and antenna or cell-tower information. These antenna-level signals are then used to automatically infer the location of devices over time. Such monitoring includes all devices from the network provider, including roaming devices. The resulting activity counts were then extrapolated by the data provider to the overall population using information on the provider's market share. Because these signals arise passively during ordinary network operation, they are not subject to self-reporting bias, unlike surveys, smartphone apps, or other opt-in monitoring systems. Telecommunication data can therefore provide broad, hyperlocal measures of urban presence. In our setting, the mobile network operator processed the raw signals internally and provided only anonymized and aggregated hourly counts at the grid-cell level, without individual identifiers or movement trajectories.

We applied standard preprocessing to the raw counts. Nine dates were excluded due to incomplete data delivery from the provider, and eight hour–date combinations with implausibly low counts were imputed from the adjacent hours within each grid cell. A small number of cells exhibited measurement artifacts in which counts accumulated at commuting peaks rather than being distributed across the hours in which they occurred. We identified these cells using a neighbor-based detection procedure (queen contiguity \cite{Bivand.2013}) and imputed them with values from their highest-count non-flagged neighbors at the same date and hour. In total, 87 of $\sim$160,000 cells (0.05\%) were corrected across the ten cities. 

Supplementary Table~\ref{tab:activity_statistics} summarizes activity levels by city. Activity levels were computed as the average number of people per cell in the respective city. Frankfurt am Main records the highest mean, median, and 75th percentile activity, followed by Munich and Berlin. Dortmund and Gelsenkirchen show the lowest activity levels across all distributional measures. Across all cities, mean activity is higher on working days than on non-working days. In addition, the mean exceeds the median in every city, indicating that activity is concentrated in a subset of more active cells.

\subsection{Weather data}
\label{sec:weather_data}
Hourly air temperature observations were obtained from Deutscher Wetterdienst (DWD; German Meteorological Service) station records\cite{DWD.2024}. For each city, the geographically nearest weather station was identified by computing the city centroid and selecting the station of minimum Euclidean distance; all assignments were manually verified by map inspection. Two heat indicators were derived from the hourly temperature series following DWD definitions. A summer day was defined as any calendar day on which the maximum air temperature reached or exceeded 25$\,^\circ$C\cite{DWD.2026.Sommertag}. A tropical night was defined as any night during which the minimum air temperature remained at or above 20$\,^\circ$C, where night was taken as 20:00–08:00 local time \cite{DWD.2026.Tropennacht}.

Supplementary Table~\ref{tab:weather_statistics} reports descriptive temperature statistics for the 10 cities in the sample. Average temperatures range from 16.7\textdegree C in Hamburg to 18.9\textdegree C in Berlin. Maximum temperatures exceed 30\textdegree C while the number of summer days above 25\textdegree C also varies across cities, with Berlin, Munich, Leipzig, Frankfurt am Main, and Stuttgart recording comparatively more hot working days than the other cities in the sample. As a diagnostic for possible confounding, we tested whether daily temperatures differed systematically between workdays and non-workdays (see Supplementary Table~\ref{tab:temp_workday_balance}). Across all cities, we found no statistically significant association between workday status and temperature. Hence, workdays were not consistently hotter or cooler than non-workdays in the observation period. Here, each day was classified as a workday or non-workday. Saturdays, Sundays, and public holidays were treated as non-workdays. State-level school holiday periods were likewise recorded for each city.

\subsection{Points-of-interest data}
\label{sec:POI data}
POI data were retrieved from OpenStreetMap (OSM) \cite{OpenStreetMapcontributors.2026} using a historical snapshot dated 13 May 2024 to ensure temporal alignment with the start of the study period. For each city, we collected a predefined set of OSM tags describing urban amenities and facilities, including food and drink establishments, retail locations, educational and civic facilities, health-related services, mobility-related amenities, recreational spaces, sports facilities, and tourism-related locations. The collected OSM tags were then harmonized and grouped into six broader analytical POI categories according to their dominant urban function: (1)~\emph{Food \& Drink}, (2)~\emph{Retail \& Shopping}, (3)~\emph{Public Service}, (4)~\emph{Green Spaces}, (5)~\emph{Sport Facility}, and (6)~\emph{Tourism \& Culture}. For example, restaurants, cafés, bars, and pubs were assigned to \emph{Food \& Drink}; shops and consumer services to \emph{Retail \& Shopping}; educational, civic, administrative, health-related, and mobility-related amenities to \emph{Public Service}; parks and playgrounds to \emph{Green Spaces}; sports facilities and swimming pools to \emph{Sport Facility}; and museums, attractions, accommodation, and other tourism-related tags to \emph{Tourism \& Culture}. The full mapping from OSM tags to analytical POI categories is reported in Supplementary Table~\ref{tab:poi_categories}.

\subsection{Regression}
To analyze whether higher temperatures are associated with a redistribution of activity across different urban environments, we estimate city-specific fixed-effects regressions at the cell-hour level. As activity is observed at the cell level rather than at the POI level, the analysis cannot identify whether individuals visit specific POIs directly. Instead, the estimates capture whether activity redistributes toward cells characterized by particular POI types under hotter temperature conditions. We model the hourly activity $y$ in grid cell $i$ at hour $h$ on day $d$, and the main explanatory variables are interactions between temperature-bin indicators and cell-level POI characteristics. The specification is given by
\begin{align}
\label{eq:regression}
y_{ihd}
&=
\sum_{b \in \mathcal{B}}
\eta_b\,\mathrm{TempBin}_{hd}^{\,b}
+
\sum_{b \in \mathcal{B}} \sum_{k \in \mathcal{K}}
\beta_{bk}
\bigl(\mathrm{TempBin}_{hd}^{\,b} \times \mathrm{POI}_{i}^{\,k}\bigr)
\nonumber \\
&\quad
+ \gamma\,\mathrm{TotalActivity}_{-i,hd}
+ \delta\,\mathrm{Workday}_{d}
+ \theta\,\mathrm{SchoolHoliday}_{d}
+ \kappa\,\mathrm{EuroPeriod}_{d}
\nonumber \\
&\quad
+ \alpha_i + \lambda_h + \omega_w + \varepsilon_{ihd},
\end{align}
where $\alpha_i$ denotes cell fixed effects, $\lambda_h$ hour-of-day fixed effects, and $\omega_w$ week fixed effects. The temperature-bin indicators distinguish between cooler and hotter conditions, while the POI variables capture the built-environment composition of each cell, measured as the number of amenities of a given type in the cell, with green spaces coded as a binary indicator. The temperature-bin main effects capture the general association between hotter conditions and activity for the baseline cell, whereas the interaction terms indicate whether cells characterized by particular POI types experience an additional increase or decrease in activity in a given temperature range. The model, therefore, does not identify visits to specific POI; rather, it estimates whether hotter conditions are associated with differential changes in activity in grid cells characterized by particular POI types while accounting for changes in overall city-wide activity. Accordingly, a positive interaction coefficient indicates that, conditional on overall city activity and fixed spatial and temporal patterns, cells with that POI characteristic attract relatively more activity in the corresponding temperature range than comparable cells in the reference temperature category.

\subsection{Robustness checks}

As a robustness check for residual spatial dependence, we re-estimate the regression using standard errors clustered over spatial groups defined by second-order queen contiguity. Under this approach, cells are assigned to broader spatial clusters that connect cells sharing an edge or corner, as well as cells connected through one additional queen-neighbor step. This accounts for the possibility that nearby cells are jointly affected by common local shocks, spillovers, or neighborhood-level activity patterns that extend beyond a single grid cell \cite{Elhorst.2014}. The robustness check leaves the regression specification and coefficient estimates of Equation~(\ref{eq:regression}) unchanged and affects only the estimated standard errors. The results can be found in Supplementary~Section~\ref{sec:robustness_regression_spatial}.

\subsection{Implementation details}
\label{sec:implementation_details}

All analyses were implemented in Python 3.11.14. Data processing used \texttt{pandas} 2.3.3, \texttt{numpy} 2.3.5, \texttt{duckdb} 1.4.3, and \texttt{scikit-learn} 1.8.0. Geospatial processing used \texttt{geopandas} 1.1.1, \texttt{Shapely} 2.1.2, \texttt{libpysal} 4.14.1, and \texttt{osmnx} 2.1.0. Statistical analyses used \texttt{pyfixest} 0.50.0, \texttt{statsmodels} 0.14.6, \texttt{scipy} 1.16.3, and \texttt{pygam} 0.12.0. Figures were created using \texttt{matplotlib} 3.10.8 and \texttt{seaborn} 0.13.2. Regression models were estimated with \texttt{pyfixest}.

\section*{Data availability}

Weather data are publicly available from Deutscher Wetterdienst (\url{https://opendata.dwd.de/climate_environment/CDC/observations_germany/climate/hourly/air_temperature/historical/}), and point-of-interest data are available from OpenStreetMap. The raw telecommunication data used in this study cannot be publicly shared. Access to raw telecommunication data used for this study can be obtained from the corresponding author upon reasonable request and a sharing agreement.  The anonymized mobile network data can be obtained from T-Systems International. The data provide anonymized and aggregated mobile network information that can be used for mobility analytics, transportation planning, and other location-based analyses (\url{https://dih.telekom.com/de/motion-data}). 

\section*{Code availability}

The code used to process the data, estimate the statistical models, and generate the figures and tables is available in the public repository: \url{https://github.com/DominiqueGeissler/urban_activity_responses_to_heat_exposure}. The repository includes scripts for data preprocessing, regression analysis, robustness checks, and visualization.

\section*{Acknowledgement}

RD acknowledges support from the Minderoo Foundation.


\newpage
\bibliography{literature}


\section*{Author contribution}
D.G. conceptualized the study, developed the methodology, performed the data preparation, formal analysis, and visualization, and wrote the manuscript. All authors contributed to the conceptualization and writing of the manuscript and approved the final manuscript.

\section*{Competing interests}
The authors declare no competing interests.

\newpage

\appendix
\renewcommand{\thefigure}{A.\arabic{figure}}
\setcounter{figure}{0}
\renewcommand{\thetable}{A.\arabic{table}}
\setcounter{table}{0}

\section{Supplementary Figures}

\subsection{Average daytime activity and shift during hot days}
\label{sec:activity_change_maps_all_cities}
\begin{figure}[H]
    \centering
    \includegraphics[width=\linewidth]{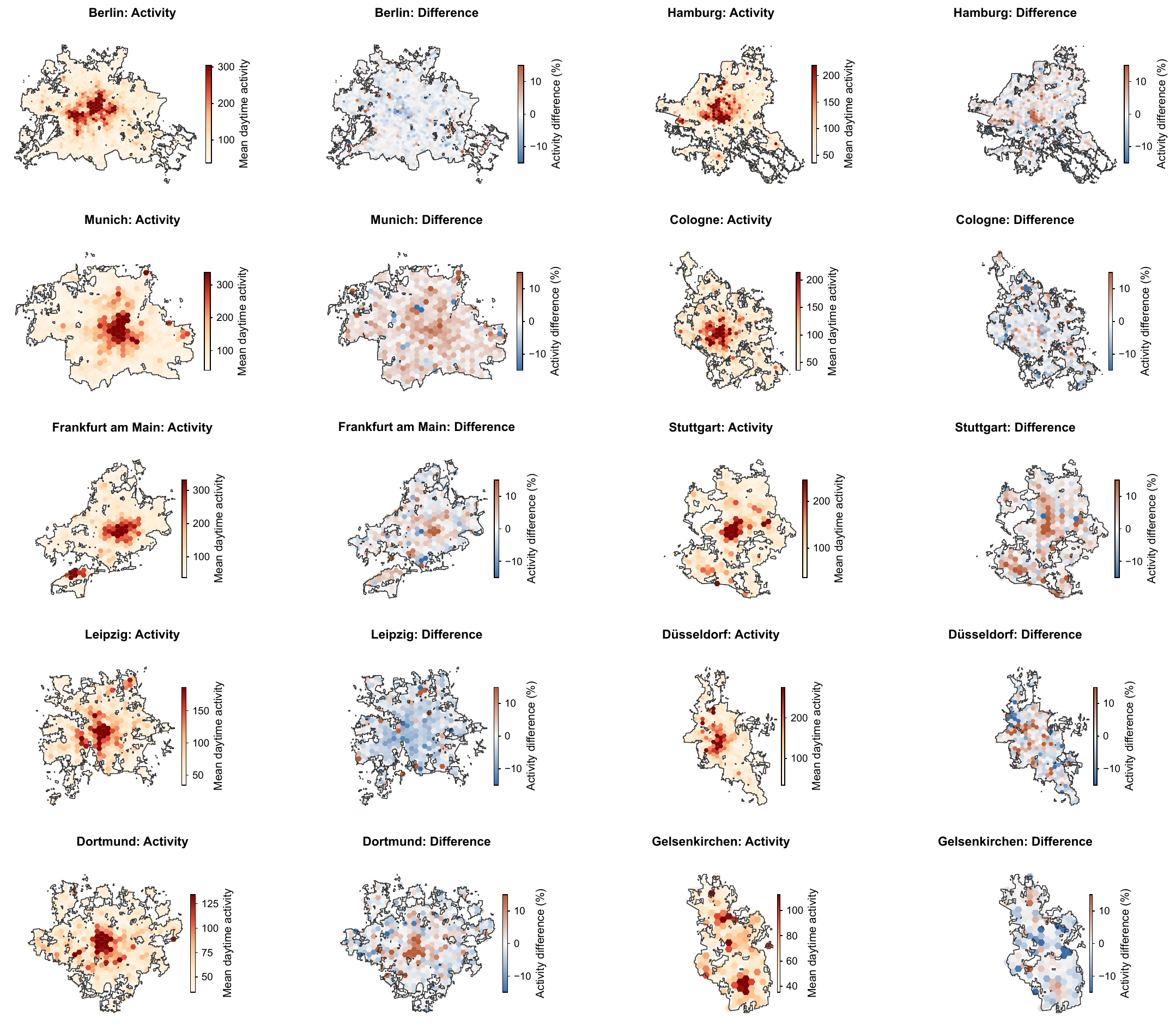}
    \caption{Average daytime activity in each city, aggregated to approximately 900 m hexagons for visualization and observed percentage difference in daytime activity on hot days, defined as days with daily maximum temperature $\geq 25\,^{\circ}$C, relative to cooler days below $25\,^{\circ}$C, using the same spatial aggregation.}
    \label{fig:activity_change_maps_all_cities}
\end{figure}

\subsection{Heterogeneity in the role of POIs across cities}
\begin{figure}[H]
    \centering
    \includegraphics[width=0.63\linewidth]{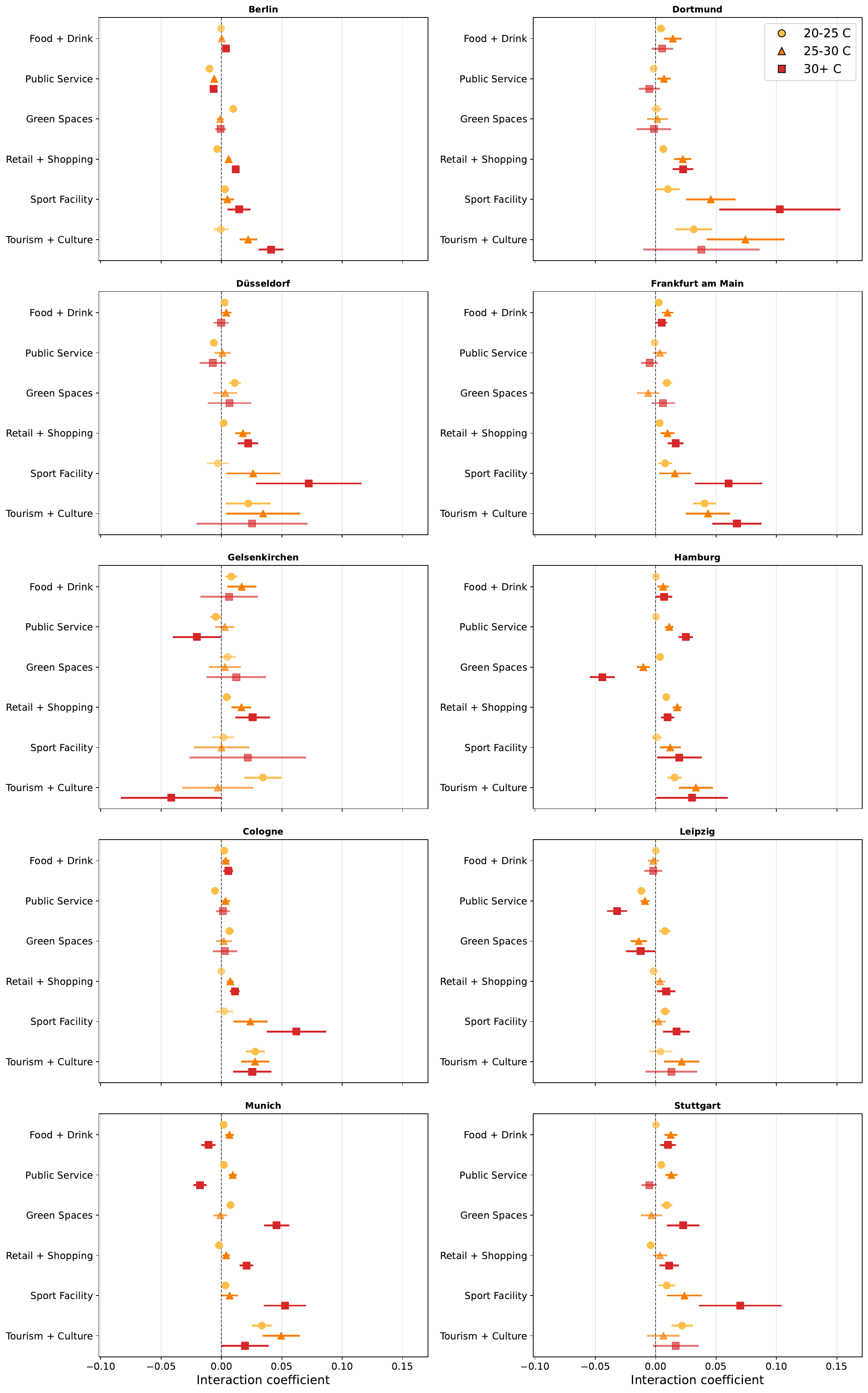}
    \caption{\textbf{Heterogeneity in the role of POIs across cities.} The figure reports estimated interaction coefficients from city-specific regressions linking temperature bins to activity near six POI categories: \emph{Food + Drink}, \emph{Retail + Shopping}, \emph{Public Service}, \emph{Green Spaces}, \emph{Sport Facility}, and \emph{Tourism + Culture}. The analysis is repeated separately for Berlin, Dortmund, Düsseldorf, Frankfurt, Gelsenkirchen, Hamburg, Köln, Leipzig, München, and Stuttgart (left to right, top to bottom). Points denote coefficient estimates and whiskers denote 95\% confidence intervals.}
    \label{fig:regression_poi}
\end{figure}

\section{Supplementary Tables}

\subsection{Population size of German cities}

\begin{table}[H]
\centering
\begin{tabular}{lr}
\toprule
City & Population \\
\midrule
Berlin & 3,755,251 \\
Hamburg & 1,892,122 \\
Munich & 1,512,491 \\
Cologne & 1,084,831 \\
Frankfurt am Main & 764,474 \\
Stuttgart & 632,865 \\
Düsseldorf & 629,047 \\
Leipzig & 619,879 \\
Dortmund & 601,402 \\
Gelsenkirchen & 264,943 \\
\midrule
Total & 11,757,305 \\
\bottomrule
\end{tabular}
\caption{Population of the 10 cities included in the analysis.}
\label{tab:city_population}
\end{table}

\clearpage

\subsection{POI categories}

\begin{table}[H]
\centering
\begin{tabular}{p{0.22\linewidth} p{0.70\linewidth}}
\toprule
\textbf{POI category} & \textbf{OpenStreetMap tag values} \\
\midrule
Food \& Drink & \texttt{restaurant}, \texttt{cafe}, \texttt{fast\_food}, \texttt{bar}, \texttt{pub}, \texttt{biergarten} \\
Retail \& Shopping & \texttt{supermarket}, \texttt{convenience}, \texttt{bakery}, \texttt{mall}, \texttt{department\_store}, \texttt{clothes}, \texttt{shoes}, \texttt{electronics}, \texttt{furniture}, \texttt{hairdresser}, \texttt{beauty}, \texttt{optician}, \texttt{chemist} \\
Public Service & \texttt{school}, \texttt{kindergarten}, \texttt{university}, \texttt{college}, \texttt{library}, \texttt{community\_centre}, \texttt{townhall}, \texttt{post\_office}, \texttt{hospital}, \texttt{clinic}, \texttt{doctor}, \texttt{doctors}, \texttt{dentist}, \texttt{pharmacy}, \texttt{physiotherapist}, \texttt{laboratory} \\
Green Spaces & \texttt{park}, \texttt{playground} \\
Sport Facility & \texttt{sports\_centre}, \texttt{fitness\_centre}, \texttt{indoor\_pool}, \texttt{outdoor\_pool} \\
Tourism \& Culture & \texttt{museum}, \texttt{gallery}, \texttt{attraction}, \texttt{stadium} \\
\bottomrule
\end{tabular}
\caption{Aggregation of OpenStreetMap point-of-interest tags into POI categories.}
\label{tab:poi_categories}
\end{table}

\clearpage

\subsection{Relationship between midday temperature and city-wide activity using raw activity counts}

\begin{table}[h]
\centering
\small
\begin{tabular}{@{}lrrr@{\hspace{.5cm}}rrr@{}}
\toprule
 & \multicolumn{3}{c}{Workday} & \multicolumn{3}{c}{Non-workday} \\
\cmidrule(lr){2-4} \cmidrule(lr){5-7}
City & Coef. & Std. error & $p$-value & Coef. & Std. error & $p$-value \\
\midrule
All cities & 50{,}718 & 35{,}612 & 0.154 & 157{,}011 & 59{,}447 & $\leq 0.01$ \\
Berlin & --58{,}566 & 21{,}972 & $\leq 0.01$ & 47{,}761 & 27{,}381 & 0.081 \\
Dortmund & --2{,}577 & 1{,}720 & 0.134 & 1{,}769 & 2{,}597 & 0.496 \\
Düsseldorf & --4{,}694 & 2{,}113 & $\leq 0.05$ & 4{,}610 & 4{,}720 & 0.329 \\
Frankfurt am Main & --10{,}346 & 7{,}019 & 0.140 & 14{,}816 & 5{,}887 & $\leq 0.05$ \\
Gelsenkirchen & --1{,}236 & 559 & $\leq 0.05$ & 734 & 523 & 0.160 \\
Hamburg & --7{,}174 & 16{,}668 & 0.667 & 21{,}862 & 14{,}207 & 0.124 \\
Cologne & --4{,}993 & 3{,}086 & 0.106 & 6{,}176 & 3{,}899 & 0.113 \\
Leipzig & --5{,}348 & 2{,}166 & $\leq 0.05$ & --4{,}500 & 1{,}704 & $\leq 0.01$ \\
Munich & 4{,}485 & 3{,}538 & 0.205 & 4{,}352 & 3{,}060 & 0.155 \\
Stuttgart & 696 & 1{,}963 & 0.723 & 1{,}563 & 1{,}757 & 0.374 \\
\bottomrule
\end{tabular}
\caption{\textbf{Estimated relationship between midday temperature and raw city-wide activity counts by city and day type.} The table reports ordinary least squares estimates from regressions of raw city-wide activity counts on midday temperature, estimated separately by city and day type. The pooled estimates indicate a stronger positive relationship on non-workdays than on workdays, while city-specific estimates show substantial heterogeneity.}
\label{tab:temperature_activity_midday_daytype_raw}
\end{table}

\clearpage

\subsection{Activity per city by temperature bin}
\scriptsize
\renewcommand{\arraystretch}{1.25}
\begin{longtable}{llp{1.5cm}p{2.5cm}p{2.5cm}p{2.5cm}}
\toprule
\textbf{City} & \textbf{Outcome} & \textbf{$<20\,^\circ$C} & \textbf{$20$--$25\,^\circ$C} & \textbf{$25$--$30\,^\circ$C} & \textbf{$30+\,^\circ$C} \\
\midrule
\endfirsthead

\toprule
\textbf{City} & \textbf{Outcome} & \textbf{$<20\,^\circ$C} & \textbf{$20$--$25\,^\circ$C} & \textbf{$25$--$30\,^\circ$C }& \textbf{$30+\,^\circ$C} \\
\midrule
\endhead

\multirow{2}{*}{Berlin}
& Per grid cell
& 121.85
& 117.70 ($\downarrow$3.4\%)
& 117.43 ($\downarrow$0.2\%)
& 118.26 ($\uparrow$0.7\%) \\
& Total (million)
& 221.6
& 207.6 ($\downarrow$6.3\%)
& 213.3 ($\uparrow$2.7\%)
& 214.1 ($\uparrow$0.4\%) \\
\midrule

\multirow{2}{*}{Dortmund}
& Per grid cell
& 73.81
& 72.60 ($\downarrow$1.6\%)
& 73.80 ($\uparrow$1.7\%)
& 72.19 ($\downarrow$2.2\%) \\
& Total (million)
& 8.6
& 8.0 ($\downarrow$6.5\%)
& 8.5 ($\uparrow$6.5\%)
& 8.0 ($\downarrow$6.8\%) \\
\midrule

\multirow{2}{*}{Düsseldorf}
& Per grid cell
& 114.14
& 112.92 ($\downarrow$1.1\%)
& 113.49 ($\uparrow$0.5\%)
& 111.31 ($\downarrow$1.9\%) \\
& Total (million)
& 13.5
& 12.7 ($\downarrow$5.6\%)
& 13.3 ($\uparrow$4.8\%)
& 12.7 ($\downarrow$4.8\%) \\
\midrule

\multirow{2}{*}{Frankfurt am Main}
& Per grid cell
& 134.94
& 133.20 ($\downarrow$1.3\%)
& 135.25 ($\uparrow$1.5\%)
& 137.91 ($\uparrow$2.0\%) \\
& Total (million)
& 40.0
& 39.4 ($\downarrow$1.5\%)
& 40.6 ($\uparrow$2.9\%)
& 41.5 ($\uparrow$2.2\%) \\
\midrule

\multirow{2}{*}{Gelsenkirchen}
& Per grid cell
& 69.55
& 68.83 ($\downarrow$1.0\%)
& 69.45 ($\uparrow$0.9\%)
& 67.65 ($\downarrow$2.6\%) \\
& Total (million)
& 3.2
& 3.1 ($\downarrow$2.3\%)
& 3.1 ($\uparrow$0.5\%)
& 2.9 ($\downarrow$7.4\%) \\
\midrule

\multirow{2}{*}{Hamburg}
& Per grid cell
& 100.69
& 94.32 ($\downarrow$6.3\%)
& 99.81 ($\uparrow$5.8\%)
& 99.09 ($\downarrow$0.7\%) \\
& Total (million)
& 68.3
& 61.9 ($\downarrow$9.4\%)
& 69.3 ($\uparrow$11.9\%)
& 69.3 ($\uparrow$0.1\%) \\
\midrule

\multirow{2}{*}{Cologne}
& Per grid cell
& 106.36
& 107.52 ($\uparrow$1.1\%)
& 104.46 ($\downarrow$2.8\%)
& 105.69 ($\uparrow$1.2\%) \\
& Total (million)
& 19.5
& 19.1 ($\downarrow$1.5\%)
& 18.9 ($\downarrow$1.1\%)
& 19.8 ($\uparrow$4.5\%) \\
\midrule

\multirow{2}{*}{Leipzig}
& Per grid cell
& 106.42
& 105.47 ($\downarrow$0.9\%)
& 99.28 ($\downarrow$5.9\%)
& 95.56 ($\downarrow$3.7\%) \\
& Total (million)
& 12.0
& 11.5 ($\downarrow$4.5\%)
& 10.6 ($\downarrow$7.5\%)
& 9.9 ($\downarrow$6.7\%) \\
\midrule

\multirow{2}{*}{Munich}
& Per grid cell
& 118.67
& 119.07 ($\uparrow$0.3\%)
& 125.09 ($\uparrow$5.1\%)
& 111.97 ($\downarrow$10.5\%) \\
& Total (million)
& 31.1
& 30.9 ($\downarrow$0.8\%)
& 34.0 ($\uparrow$10.0\%)
& 29.8 ($\downarrow$12.3\%) \\
\midrule

\multirow{2}{*}{Stuttgart}
& Per grid cell
& 99.53
& 93.19 ($\downarrow$6.4\%)
& 103.32 ($\uparrow$10.9\%)
& 107.00 ($\uparrow$3.6\%) \\
& Total (million)
& 11.2
& 10.4 ($\downarrow$7.1\%)
& 12.4 ($\uparrow$18.4\%)
& 12.8 ($\uparrow$3.4\%) \\
\bottomrule
\caption{\textbf{City-specific median activity by temperature bin.} Each cell reports the median activity in the respective daily maximum temperature bin. From the second bin onward, brackets report the percentage change in the median relative to the previous temperature bin within the same city and outcome. Total city activity is reported in millions of activity counts.}
\label{tab:temperature_bin_activity_compact_median}\\
\end{longtable}

\clearpage
\normalsize
\subsection{Consecutive hot days by city}

\begin{table}[H]
\centering
\begin{tabular}{lrrrr}
\toprule
City & Periods & Hot-period days & Duration range & Mean temperature \\
\midrule
Berlin & 3 & 16 & 4--7 & 27.4\,$\,^\circ$C \\
Hamburg & 3 & 11 & 3--5 & 27.6\,$\,^\circ$C \\
Munich & 3 & 15 & 4--7 & 28.0\,$\,^\circ$C \\
Cologne & 2 & 8 & 3--5 & 28.3\,$\,^\circ$C \\
Frankfurt am Main & 2 & 11 & 5--6 & 28.8\,$\,^\circ$C \\
Stuttgart & 1 & 6 & 6 & 28.0\,$\,^\circ$C \\
Leipzig & 2 & 10 & 4--6 & 29.0\,$\,^\circ$C \\
Düsseldorf & 1 & 4 & 4 & 29.0\,$\,^\circ$C \\
Dortmund & 1 & 4 & 4 & 28.6\,$\,^\circ$C \\
Gelsenkirchen & 1 & 4 & 4 & 28.5\,$\,^\circ$C \\
\bottomrule
\end{tabular}
\caption{Sustained hot periods by city. Sustained hot periods are defined as at least three consecutive days with daily maximum temperature $\geq 25\,^\circ$C.}
\label{tab:consecutive_hot_days}
\end{table}

\clearpage

\subsection{Activity levels by city}

\begin{table}[H]
\centering
\small
\begin{tabular}{lrrrrrrr}
\toprule
City & $N$ Cells & Mean & 25th perc. & Median & 75th perc. & Workday & Non-Workday \\
\midrule
Berlin & 43,829 & 116.33 & 47.00 & 75.00 & 137.00 & 119.71 & 108.81 \\
Dortmund & 11,353 & 72.88 & 39.00 & 53.00 & 80.00 & 73.78 & 71.42 \\
Düsseldorf & 9,087 & 111.62 & 46.00 & 73.00 & 130.00 & 115.15 & 105.75 \\
Frankfurt am Main & 10,432 & 131.72 & 52.00 & 83.00 & 146.00 & 137.06 & 119.44 \\
Gelsenkirchen & 4,898 & 69.52 & 40.00 & 54.00 & 80.00 & 70.04 & 68.67 \\
Hamburg & 29,800 & 96.40 & 43.00 & 63.00 & 106.00 & 99.38 & 89.52 \\
Cologne & 15,537 & 105.37 & 46.00 & 70.00 & 121.00 & 107.47 & 101.85 \\
Leipzig & 10,496 & 102.98 & 45.00 & 71.00 & 124.00 & 107.41 & 99.20 \\
Munich & 15,668 & 119.00 & 53.00 & 80.00 & 134.00 & 124.43 & 107.76 \\
Stuttgart & 8,846 & 99.28 & 45.00 & 66.00 & 112.00 & 103.87 & 89.36 \\

\bottomrule
\end{tabular}
\caption{Descriptive statistics of activity levels by city (per grid cell average).}
\label{tab:activity_statistics}
\end{table}

\clearpage
\subsection{Temperature statistics for each city}

\begin{table}[H]
\centering
\small
\begin{tabular}{lccccc}
\toprule
City & Min ($\,^\circ$C) & Max ($\,^\circ$C) & Avg ($\,^\circ$C) & Avg Workdays ($\,^\circ$C) & Avg Non-workdays ($\,^\circ$C) \\
\midrule
Berlin & 9.2 & 31.5 & 18.9 & 23.1 & 23.9 \\
Dortmund & 6.1 & 31.3 & 16.8 & 21.1 & 21.4 \\
Düsseldorf & 7.1 & 31.0 & 16.9 & 20.9 & 21.0 \\
Frankfurt am Main & 10.0 & 32.6 & 18.1 & 22.6 & 22.1 \\
Gelsenkirchen & 7.8 & 30.7 & 16.9 & 21.0 & 21.0 \\
Hamburg & 6.5 & 31.4 & 16.7 & 20.8 & 20.7 \\
Cologne & 8.0 & 31.7 & 17.5 & 21.5 & 21.6 \\
Leipzig & 7.7 & 31.7 & 18.0 & 22.6 & 23.2 \\
Munich & 9.0 & 33.3 & 17.8 & 22.5 & 22.2 \\
Stuttgart & 8.8 & 31.8 & 17.6 & 21.7 & 21.7 \\
\bottomrule
\end{tabular}
\caption{Descriptive statistics of temperature by city.}
\label{tab:weather_statistics}
\end{table}

\clearpage

\subsection{Temperature differences between workdays and non-workdays}

\begin{table}[htbp]
\centering
\begin{tabular}{lccc}
\toprule
City & Workday coefficient & Std. error & $p$-value \\
\midrule
Berlin              & --1.420 & 1.086 & 0.197 \\
Dortmund            & --0.371 & 1.056 & 0.727 \\
Düsseldorf          & --0.269 & 1.059 & 0.801 \\
Frankfurt am Main   &  0.389 & 1.070 & 0.718 \\
Gelsenkirchen       & --0.215 & 1.066 & 0.841 \\
Hamburg             & --0.046 & 1.118 & 0.968 \\
Köln                & --0.305 & 1.104 & 0.784 \\
Leipzig             & --1.092 & 1.102 & 0.326 \\
München             &  0.306 & 1.282 & 0.813 \\
Stuttgart           & --0.036 & 1.192 & 0.976 \\
\bottomrule
\end{tabular}
\caption{\textbf{Daily temperature differences between workdays and non-workdays.} Each row reports the coefficient from a city-specific OLS regression of daily temperature on a workday indicator. Negative coefficients indicate cooler workdays. None of the coefficients are statistically significant at conventional levels (all $p > 0.05$).}
\label{tab:temp_workday_balance}
\end{table}

\clearpage

\section{Supplementary Methods}

\subsection{Robustness check for residual spatial dependence}
\label{sec:robustness_regression_spatial}

\begin{figure}
    \centering
    \includegraphics[width=0.7\linewidth]{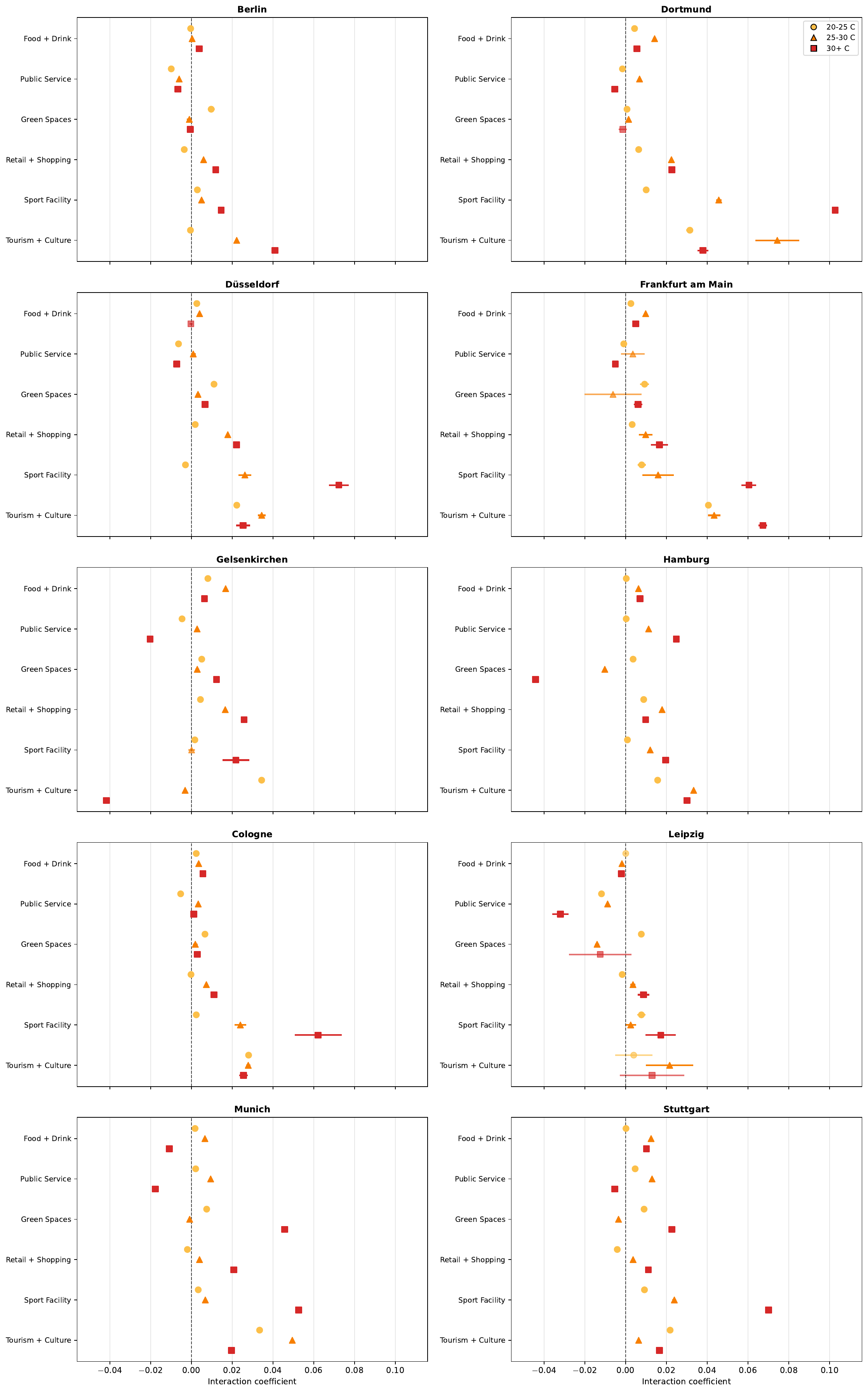}
    \caption{Heterogeneity in the role of POIs across cities using standard errors clustered over neighbors}
    \label{fig:robustness_regression_spatial_dependence}
\end{figure}

We re-estimate the activity redistribution of people towards urban environments with standard errors clustered over spatial groups defined by second-order queen contiguity as a robustness check for residual spatial dependence (see Figure~\ref{fig:robustness_regression_spatial_dependence}). We find robust results for the relationship between activity redistribution, temperature, and POIs. This indicates that the main findings are not driven by unmodeled local spatial dependence.

\end{document}